%% file: main.tex
\documentclass[a4paper,11pt]{article}
\usepackage{jinstpub}
\usepackage{lineno}
\usepackage{xcolor}
\usepackage{soul}

\newcommand{\iso}[2]{\mbox{$^{#1}$#2}}
\newcommand{\us}{\mbox{$\mu$s}}

\begin{document}

\title{Predicting the single-site and multi-site event discrimination power of dual-phase time projection chambers}

\author[a]{A.B.M.R. Sazzad}
\author[b]{C.A. Hardy } 
\author[c]{X. Dai} 
\author[a]{J. Xu } 
\author[d]{B. Lenardo} 
\author[a]{F. Sutanto} 
\author[c]{N. Antipa} 
\author[a]{J. Koertzen} 
\author[e]{P. John} 
\author[a]{A. Akinin} 
\author[a]{T. Pershing} 
\affiliation[a]{Lawrence Livermore National Laboratory, 7000 East Ave., Livermore, CA 94550, USA}
\affiliation[b]{Physics Department, Stanford University, Stanford, CA 94305, USA}
\affiliation[c]{Department of Electrical and Computer Engineering, University of California San Diego, La Jolla, CA 92093-0112, USA}
\affiliation[d]{SLAC National Accelerator Laboratory, Menlo Park, CA 94025, USA}
\affiliation[e]{Department of Computer Science and Engineering, Washington University in St. Louis, St. Louis, MO 63130-489, USA}



\emailAdd{xu12@llnl.gov}

\abstract{Dual-phase xenon time projection chambers (TPCs) are widely used in searches for rare dark matter and neutrino interactions, in part because of their excellent position reconstruction capability in 3D. Despite their millimeter-scale resolution along the charge drift axis, xenon TPCs face challenges in resolving single-site (SS) and multi-site (MS) interactions in the transverse plane. In this paper, we build a generic TPC model with an idealized signal readout, and use Fisher Information (FI) to predict its theoretical capability of differentiating SS and MS events using the electroluminescence signal. We also demonstrate via simulation that, when only statistical photon noise is present, the theoretical limits can be approached with conventional reconstruction algorithms like maximum likelihood estimation, and with a convolutional neural network classifier. The implications of this study on future TPC experiments will be discussed.  
}

\begin{NoHyper}
\maketitle
\end{NoHyper}


\section{Introduction}
\label{sec:intro}

The time projection chamber (TPC)~\cite{Nygren1974_TPC} is one of the most sensitive particle detection technologies. A TPC is a charge drift chamber that transports ionization electrons produced by particle interaction across the detector volume with an electric field; at the end of the detector the electrons are collected with a position sensitive charge sensing device. 
The charge sensor provides two-dimensional (2D) position information in the collection plane perpendicular to the electric field. 
In addition, if the interaction time is known, for example, via an accelerator beam signal or prompt scintillation accompanying ionization production, the third coordinate can be reconstructed from the electron drift time, enabling full three-dimensional (3D) event reconstruction. 
For gas TPCs that produce relatively long particle tracks, the track ionization density can be measured, allowing the particle type to be determined. For scintillating TPCs, the charge-to-scillation ratio may also help identify different particle interactions~\cite{Xenon10_ERNR}. In addition, combining the scintillation and ionization energy channels enables an accurate energy reconstruction for the interaction 
by reducing the inter-channel fluctuation~\cite{Aprile2007_YieldCorrelation}. 

TPCs using liquid argon and xenon, which combine the benefits of a bright scintillator, a large target mass, and low electron affinity, are widely used in the search for rare particle interactions such as those expected from neutrinos and dark matter~\cite{LZ_detector, LUX_2016DMresults, nEXOPDP, XENON_SI-DM_2018, PandaX-4T_SI-DM_2021,Darkside20kPDP}. 
A dual-phase noble liquid TPC further enhances the TPC's sensitivity to low-energy interactions by introducing a thin gas layer above a liquid target. 
When ionization electrons are drifted into the gas, they can produce a large number of electroluminescence photons under a strong electric field. Therefore, both scintillation and ionization signals can be conveniently collected with a set of photosensors. Typically, an extracted electron can produce hundreds of electroluminescence photons, allowing signals as small as single ionization electrons to be detected with high efficiency~\cite{XENON10_SubGeVDM}. 
Dual-phase argon and xenon TPCs have been widely used in dark matter search experiments that primarily focus on very low-energy signals. Currently, there are three xenon TPCs with multi-ton target masses in operation, including the LZ experiment~\cite{LZ_detector} at the Sanford Underground Research Facility (SURF), the XENONnT experiment~\cite{xenonnt_dmsearch} at Laboratori Nazionali del Gran Sasso (LNGS), Italy, and the PandaX-nT experiment~\cite{PandaX-4T_SI-DM_2021} at the Jinping Underground Laboratory in China. 
These detectors have been leading the searches for a range of hypothetical dark matter interactions, and their successes have inspired even larger detectors such as XLZD~\cite{XLZD2024_DesignBook} and PandaX-20T to be proposed.

Despite their 3D reconstruction capabilities, dual-phase TPCs exhibit asymmetric position resolution. In the direction of the electric field (usually referred to as the $z$ coordinate), a typical position sensitivity of $\mathcal{O}$(1)~mm can be achieved because of the $\mathcal{O}$(1)~mm/\us\ drift speed of electrons in liquid argon and xenon. Interactions with multiple vertices separated by a few millimeters in $z$ result in multiple detected ionization pulses or piled-up pulses with distorted shapes. These events could be separated from single-site (SS) events in the analysis. 
In directions perpendicular to the electric field ($x-y$ plane), the position information is encoded in the photon hit pattern in the photosensor array next to the gas electroluminescence volume. The isotropic nature of the photon emission causes the light to be broadly distributed among photosensors, so the detector's ability in reconstructing the $x-y$ position is highly dependent on detector configurations such as the size of photosensors and their separation from the electroluminescence region. Although mm-scale $x-y$ resolution has been demonstrated for high-energy SS events with cm-scale photosensors~\cite{LUX2018_PositionReconstruction}, the ability of a dual-phase TPC with optical readouts to resolve multi-site (MS) events in the $x-y$ plane has not been studied. 

One important application of SS/MS discrimination is in the search for neutrino-less double beta decays ($0\nu\beta\beta$) of the isotope $^{136}$Xe in a dual-phase xenon TPC~\cite{nEXO_sensitivity_2018}. $0\nu\beta\beta$ is predicted to occur in some double beta decay isotopes if neutrinos are Majorana fermions. A $^{136}$Xe $0\nu\beta\beta$ signal consists of two $\beta$ electrons whose energies sum to exactly $Q_{\beta\beta} = 2.457$~MeV, while the primary background is induced by $\gamma$ rays from trace amounts of natural radioactivity in the detector materials. In liquid xenon, MeV-electrons are stopped within millimeters and the events usually appear point-like, while gamma rays tend to Compton scatter and deposit energy in multiple locations separated by several centimeters. Reconstruction of event topology therefore provides a powerful means of background discrimination. The nEXO experiment proposes to search for $0\nu\beta\beta$ in $^{136}$Xe with $\sim$5 tons of enriched xenon in a  single-phase TPC and is expected to fully cover the inverted hierarchy region of neutrino mass with 10 years of operation~\cite{nEXO_sensitivity_2018}. Because of its large total mass (60--80 ton), the dual-phase TPC of XLZD would contain a comparable mass of $^{136}$Xe to that of nEXO and its sub-percent energy resolution near the $^{136}$Xe $Q$-value enables it to carry out a sensitive $0\nu\beta\beta$ search~\cite{XLZD2025_0vbb}. In both cases, $^{214}$Bi decay gamma rays are expected to be the dominant background in $0\nu\beta\beta$ energy region even with meticulous material cleanliness protocols~\cite{nEXO_sensitivity_2018,XLZD2025_0vbb}. Therefore, a thorough understanding of the dual-phase xenon TPC's capability in separating SS and MS events is critical. 

Dark matter searches utilizing dual-phase TPCs can also benefit from improved SS/MS discrimination. For example, xenon TPCs rely heavily on the scintillation-ionization signal ratio to differentiate nuclear recoils from electron recoils, the calibration of which, however, could be skewed by contamination from MS backgrounds by neutrons and gammas~\cite{LZ_backgrounds}. 
Moreover, in searches for Migdal interactions in liquid xenon, which predict enhanced signals for low-energy nuclear recoils and can substantially boost the low-mass dark matter sensitivity of existing experiments, multi-scatter neutron interactions are found to be a leading background source~\cite{Migdal_search_Xu2024}. Thus, SS/MS discrimination is broadly applicable across rare-event searches. 

This paper is organized as follows. Section.~\ref{sec:theory} describes a simplified dual-phase TPC model, and predicts its sensitivity in identifying MS events using Fisher Information (FI) theory. Section~\ref{sec:methods} compares traditional and machine learning reconstruction algorithms against these theoretical expectations. 
The implications of this study on rare neutrino and dark matter searches and their optimization will be discussed in Sec.~\ref{sec:disc}, and conclusions are given in Sec.~\ref{sec:conclusion}.

\section{Detector model and resolution predictions}
\label{sec:theory}

We develop a generic dual-phase TPC model with a light-based readout of ionization signals to study its theoretical performance in reconstructing SS and MS events across different event topologies as a function of the photosensor size and the total detected photon count. 
Photosensor diameters of 3", 2", 1", 15~mm, and 6~mm are considered, which represent the sizes of commonly used PMTs and SiPMs in dual-phase TPCs. 
Poissonian photon fluctuations are accounted for in this theoretical study but instrumentation effects that may introduce nonlinear signal response or additional noise are not considered. 
The detector model assumes direct photon detection with photonsensors, so the obtained results do not apply to liquid argon TPCs that use surface wavelength shifter coatings such as DarkSide50~\cite{DarkSide_S2only_2018}, although an argon TPC model can be constructed and studied using a similar method. 

\subsection{Detector model}
\label{sec:detector}

\begin{figure}[!ht]
    \centering
    \includegraphics[width=0.48\textwidth]{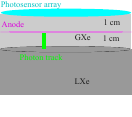}
    \includegraphics[width=0.5\textwidth]{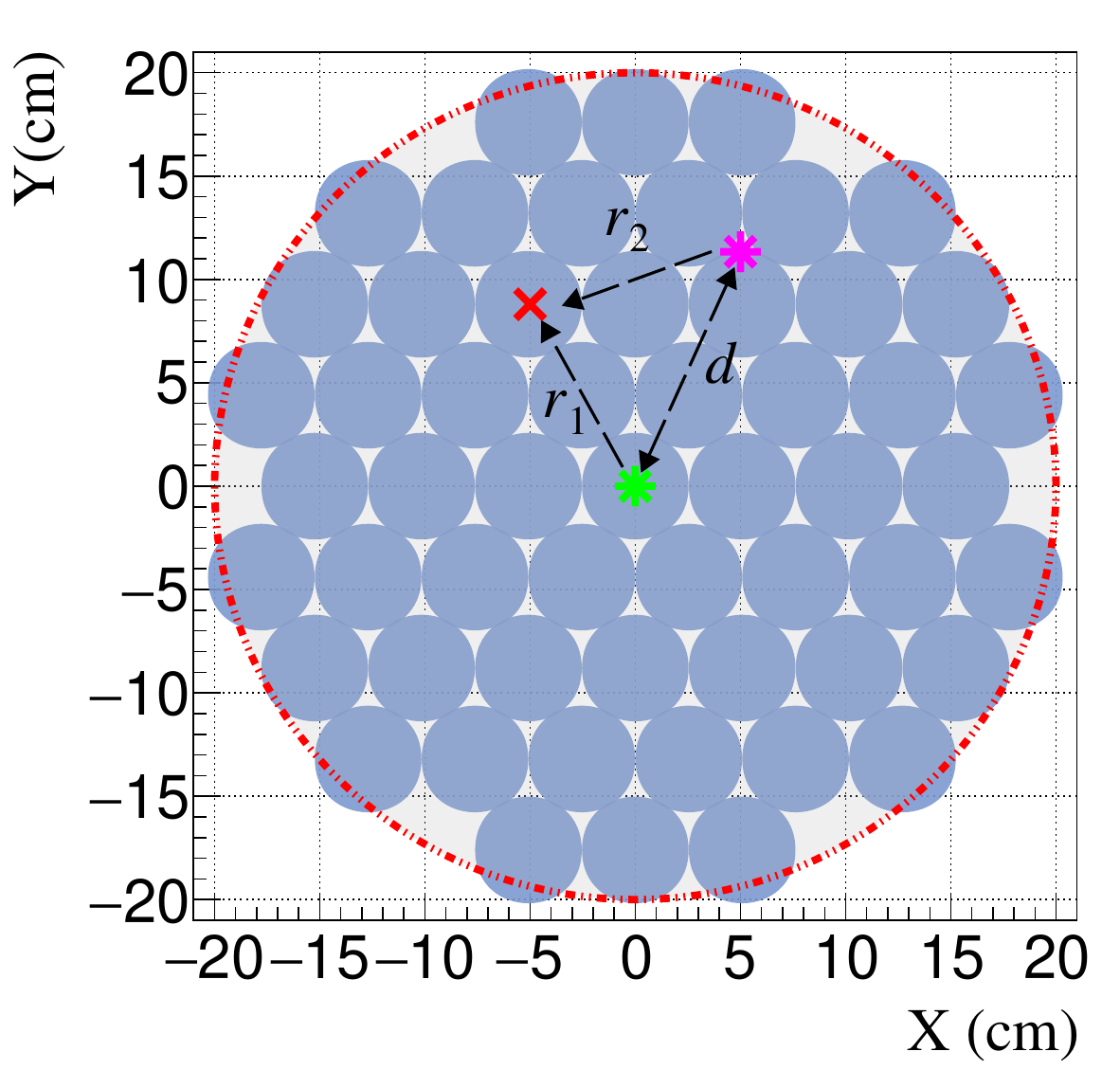}
    \caption{\textbf{Left} - an illustration of the generic TPC model used in our studies; \textbf{Right} - a cross-section view of a 40~cm TPC illustrating the photosensor array configuration; the asterisks represent two light emission sites separated by a distance $d$ and $r_1$ and $r_2$ represent the radial distances between the light sources and the PMT under investigation.}
    \label{fig:setup}
\end{figure}

The detector model, as illustrated in Fig.~\ref{fig:setup}, is simplified to capture the essential features of a  dual-phase xenon TPC.
The electroluminescence region is modeled as a 1~cm xenon gas layer above a liquid–gas interface, where the electric field is assumed to be perfectly perpendicular to the liquid surface. With this approximation, electroluminescence photons excited by extracted electrons at a certain $x-y$ location are produced uniformly along the vertical drift paths in this gap. Our model omits physical TPC electrodes, which generate the TPC drift field but may introduce field non-uniformities near the grid surfaces and cause shadowing effect in the light distribution. 
An array of photonsensors is placed 2~cm above the liquid surface to create a 1~cm separation between the upper end of photon tracks and the nearest sensor. 
This light source-sensor separation is smaller than those used in large TPCs, but is chosen to allow the studies of small sensors with the same TPC configuration; the effects of the sensor separation are discussed in Secs.~\ref{sec:lux} and \ref{sec:optimize}. 
Each sensor is assumed to be circular in shape~\footnote{SiPMs are often fabricated in square shapes, which can be modeled using the same method as described in this work; we model all sensors in the same shape to directly compare their performances. } in a hexagonal close-packed configuration, yielding a photocathode fill factor of $\sim$91\%. 
The sensor array is assumed to be sufficiently large to capture essentially all photons that could be collected, simulating the scenario of near-center events in a large TPC. While partial photon collection at detector edges could be implemented for a specific TPC geometry, such effects are not considered in this theoretical study. 
In practice, significant Fresnel reflections at large incident angles, combined with the small vertical separation between light source and sensor, limits the acceptance of photons from large radial distances. When photons are emitted from the center of the TPC, a 40~cm sensor array is able to capture virtually all detectable photons. Studies requiring light production far away from the center use larger detector sizes accordingly. 

A Geant4-based Monte Carlo simulation of the optical process is used to predict the photon detection pattern in the photosensor array for electroluminescence light emitted from a specific $x-y$ location. 
In the simulation, all material surfaces, except the liquid-gas boundary, are assumed to be photon-absorbing so that no reflectance of light is in effect except at the liquid surface. 
Despite the simplification, our simulation approximately reproduces the radial photon distribution in the LUX experiment~\cite{LUX2018_PositionReconstruction} when the same sensor placement is used (Sec.~\ref{sec:lux}). 
The photons are produced isotropically from a vertical track at their assigned $x-y$ locations, spanning between the liquid surface and the nominal anode. Geant4 optical physics is used to track the photons until they are absorbed by the photosoensors or other non-sensitive surfaces. 
For convenience we assume a sensor quantum efficiency (QE) of 1, but we will report all results as a function of detected photon number, so a $<$1 QE only modestly affects the results by introducing an additional fluctuation in the event-by-event photon count in a given sensor. 
Similarly, saturation effects and instrumentation noise are not included in this study. 

\begin{figure}[!ht]
    \centering
    \includegraphics[width=0.99\textwidth]{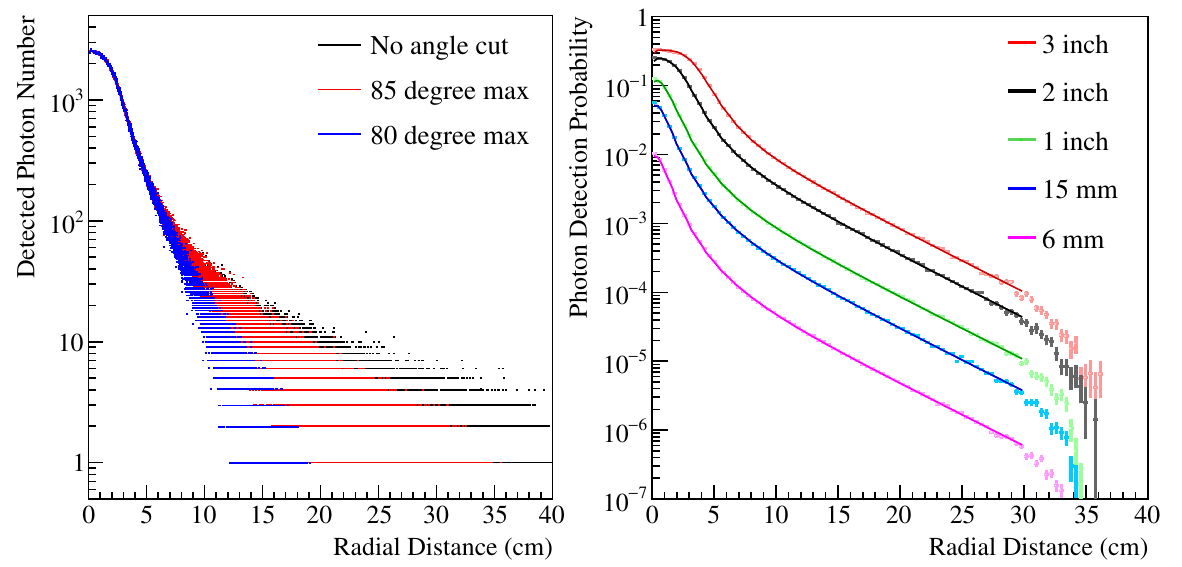}
    \caption{\textbf{Left} - the simulated number of photons detected by 2" PMTs, as a function of the radial distance between the PMT center and the light source, for every 10,000 emitted photons; different colors represent different maximum photon acceptance angles (black-90 degree, red-85 degree, blue-80 degree); \textbf{Right} - mean photon detection probability as a function of radial distance (85 degree cutoff), along with their best fit functions, for individual sensors of different sizes (from top to bottom: 3", 2", 1", 15~mm and 6~mm).}
    \label{fig:LRF}
\end{figure}

Because of the symmetry provided by the circular shape of the photosensors and their hexagonal packing, the probability for a photon to be detected by a sensor is mostly determined by the horizontal distance between the photon source and the sensor center ($r$). 
Figure.~\ref{fig:LRF} (left) shows the number of photons detected by 2" sensors, as a function of $r$, when 10,000 photons are emitted from the source. 
The probability decreases monotonically due to the inverse square law for photon flux, the decreasing solid angle subtended by the sensor, and the angular acceptance at the sensor surface. 
For small $r$ values, the diameter of the sensor dominates the aforementioned effects, with a relatively mild $r$ dependence within a sensor radius, beyond which a sharp drop in probability is observed. 
The different data sets correspond to different angular acceptance assumptions, which primarily impact photons hitting faraway sensors at large incident angles. 
For the following work, we choose a maximum photon acceptance angle of 85 degrees, motivated by the $\sim$50\% Fresnel reflectance of quartz at this angle. 

Hereafter, we approximate the simulated photon detection probability as a function of the radial distance $r$, referred to as the light response function (LRF), with an analytical form: 
\begin{equation}
    \eta(r) = A \cdot e^{\left(-a \cdot \frac{\rho}{1+\rho^{1-\alpha}} - \frac{b}{1+\rho^{-\alpha}}\right)}, \; \rho=\frac{r}{r_0} 
    \label{eq:lrf}
\end{equation}
This parameterization is adopted from~\cite{posReconstruct_Solovov2012}, and is observed to describe our simulation result well, especially at large $r$ values. 
The $r$ profiles of the LRF for 3", 2", 1", 15~mm, and 6~mm sensor sizes, together with their best-fit functions using Eq.~\ref{eq:lrf}, are shown in Fig.~\ref{fig:LRF} (right). 

In the following studies, we only use the analytical form of the LRF for both theoretical studies and reconstruction demonstrations for the following reasons. First, this approach eliminates the need for the sensor hit pattern to be simulated for each photon collection scenario through the slow Monte Carlo process; in some cases, such a simplification is necessary to enable the theoretical study. 
Second, even high-statistics Monte Carlo simulations will have non-negligible fluctuations for edge cases where the photon collection efficiency is low, and such fluctuations may bias the theoretical predictions. 
Finally, in the reconstruction studies, by simulating photon hit patterns and reconstructing event positions using the same LRF, we directly compare the reconstruction performance with theoretical predictions with little ambiguity from systematic uncertainties. 
However, we note that this approach assumes that all physics is known to the reconstruction, so the result may be better than what can be realistically achievable in an experiment.

\subsection{Fisher Information and Cramer-Rao Lower Bound}
\label{sec:fisher}

Fisher Information (FI) is a likelihood-based method to quantify the information contained in measured data about parameters of interest. 
The gradient of the log-likelihood function, which is also called the score, measures the change in the log-likelihood value with respect to a small change in a chosen parameter:
\begin{equation}
    s(\Theta) = \frac{\partial}{\partial \theta} \log \mathcal{L}(X | \Theta)
\end{equation}
where  $X=\{x_k\}$ represents the measured data and $\Theta=\{\theta_i\}$ are the parameters of interest. For a parameter that is meaningfully constrained by measured data, the expectation of the score over all possible measurement results should be 0 at the true parameter value, which is the basis for the maximum likelihood estimation method.  
The FI matrix can be constructed from the covariance of the score vector. 
\begin{equation}
        I(\Theta) = E \left[ \left(\frac{\partial}{\partial \theta} \log \mathcal{L}(X | \Theta)\right)^T \left(\frac{\partial}{\partial \theta} \log \mathcal{L}(X | \Theta)\right) \right]
    \label{eq:fi}
\end{equation}
where the expectation is calculated over the observable space $X$. Under some regularity conditions, the FI matrix may also be equivalently constructed as
\begin{equation}
        I(\Theta) = - E \left[ \left(\frac{\partial}{\partial \theta} \right)^T \left(\frac{\partial}{\partial \theta} \right) \log \mathcal{L}(X | \Theta) \right]
    \label{eq:fi}
\end{equation}

The Cramer-Rao inequality states that the covariance of $\hat\theta_i$, which represents an unbiased estimator for the parameter $\theta_i\in\Theta$, cannot be smaller than the inverse of the FI matrix.
\begin{equation}
    Cov(\hat\theta_i) \ge [I^{-1}(\Theta)]_{ii} \label{eq:crlb}
\end{equation}
When there is a single parameter to be estimated, Eq.~\ref{eq:crlb} simply means that the smallest variance on the obtained parameter value is the inverse of the second derivative of the log-likelihood. Therefore, $\sqrt{[I^{-1}(\Theta)]_{ii}}$, or the Cramer-Rao Lower Bound (CRLB), is a limit of the precision that could be achieved with an unbiased estimator for $\hat{\theta_i}$ given the system response. 
In practice, maximizing the likelihood value, or minimizing the chi-square value or other metrics, for a given data set can provide an estimate of the parameters of interest. However, different estimators carry different uncertainties on the parameter values obtained, which may also depend on the specific implementation and initialization of the estimator. 
The smallest uncertainty in $\hat\theta$ from any unbiased estimator can be estimated by constructing and inverting the FI matrix. 

For xenon TPCs, the observables are the photon counts in each photosensor in the array, which can be modeled with a Poisson distribution. In this case, the FI matrix becomes 
\begin{align}
    I(\Theta) & = \sum_{k=1}^{K} \left( \frac{\partial \eta_k(\Theta)}{\partial \theta} \right)^T \left( \frac{\partial \eta_k(\Theta)}{\partial \theta} \right) \frac{1}{\eta_k(\Theta)} \nonumber \\
    [I(\Theta)]_{i,j} & = \sum_{k=1}^{K} \left( \frac{\partial \eta_k(\Theta)}{\partial \theta_i} \right) \left( \frac{\partial \eta_k(\Theta)}{\partial \theta_j} \right) \frac{1}{\eta_k(\Theta)}
    \label{eq:fim}
\end{align}
where $\eta_k(\Theta)$ is the Poissonian mean in the $k$th photosensor. 
As discussed in Sec.~\ref{sec:detector}, we assume that the Poissonian mean in photosensor $k$ is completely determined by the number of photons emitted and the lateral distance between the light source and the sensor center, with a formula following Eq.~\ref{eq:lrf} and parameters determined by optical simulations (Fig.~\ref{fig:LRF}). 
In our model, we ignore dark counts and read noise from the photosensors, which is justified by the capability of PMTs and SiPMs to definitively detect single photons and by their low dark count rates at cryogenic temperatures. 

For the case of SS events in our TPC model, the parameters of interest are $\Theta=\{m_1,x_1,y_1\}$, corresponding to the number of photons emitted and the $x$ and $y$ coordinates of the event, and the observables are the number of photons detected by each photosensor $X=\{n_k\}$. 
Equations~\ref{eq:lrf} and \ref{eq:fim} can then be used to compute the FI matrix, and the CRLB values on these parameters can be derived accordingly. 
For the study of double-site events (MS with 2 vertices, dubbed MS2), we add three similar parameters to describe the number of photons ($m_2$) and $x-y$  location for the second vertex. However, to conveniently evaluate the detector's sensitivity to the distance between the two vertices, instead of using the $x-y$ coordinates of the second vertex, we use its distance from the first vertex ($d$) and the relative azimuth angle ($\phi$). Therefore, the additional parameters become $\{m_2,d,\phi\}$ so that the CRLB on $d$ can be directly evaluated. 
In addition, for small $d$ values, a strong degeneracy between the photon counts at the two vertices, $m_1$ and $m_2$, is present; i.e., a decrease in $m_1$ can be approximately compensated for by increasing $m_2$ by a similar amount. To mitigate this issue, the final parameter set of $\Theta=\{m,x_1,y_1,f_2,d,\phi\}$ is used, with $m=m_1+m_2$ and $f_2=m_2/m$. 

This theoretical work is motivated by the studies of the ``resolution measure'' in optical microscopy, which investigates the smallest resolvable separation between two lit objects~\cite{FI_single_molecule_microscopy_2016}; a similar study for microscopy can be found in~\cite{Ram2006_ResolutionMeasure}.  
For interested readers, we provide a detailed derivation of the FI matrix for independent Poisson distributions with binned data in Appendix~\ref{sec:emlfi}. 

\subsection{Position resolution predictions}
\label{sec:sensitivity}

\begin{figure}[!ht]
    \centering
    \includegraphics[width=0.99\textwidth]{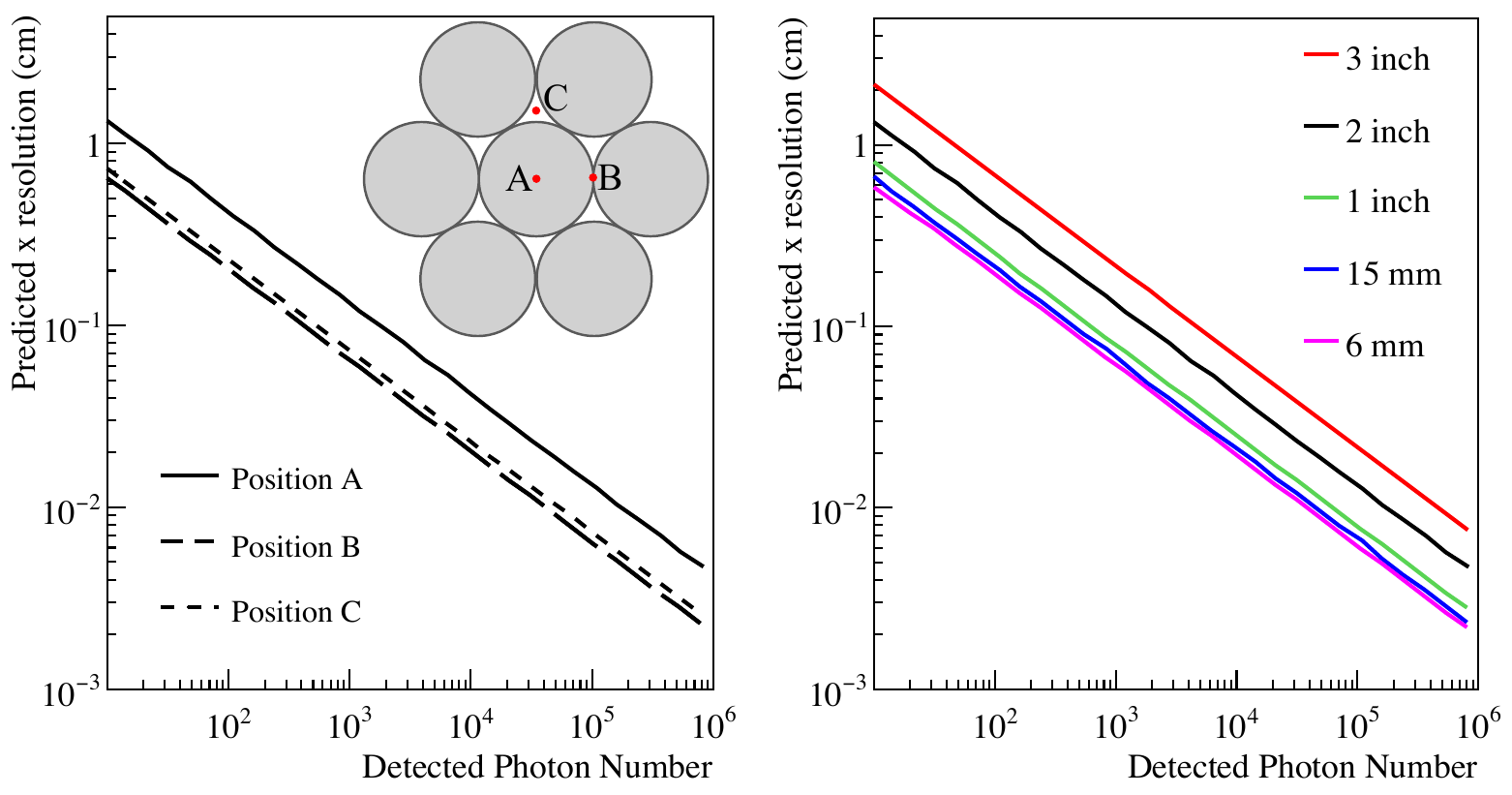}
    \caption{Predicted $x$ position reconstruction accuracy limits for SS events in our TPC model using FI and CRLB. \textbf{Left -} position resolution limits estimated for 2" PMTs, as a function of the detected photon number, at event positions A (solid), B (dashed) and C (dotted); \textbf{Right-} position resolution limits estimated for 3", 2", 1", 15~mm, and 6~mm circular photosensors (from top to bottom) for events at position A.}
    \label{fig:sslimit}
\end{figure}

For SS events, there are only 3 parameters of interest: $m_1,x_1,y_1$. 
We now use the method described in Sec.~\ref{sec:fisher} to calculate the theoretical limit on the reconstruction accuracy for each parameter that could be achieved with an unbiased reconstruction method under a range of detector configurations. 
Figure~\ref{fig:sslimit} (left) shows the derived resolution limit for $x_1$ reconstruction using 2" PMTs as a function of detected photon number for event location A (solid line), B (dashed line), and C (dotted line). 
For events at position A, the light is emitted directly under a PMT so a large fraction of light is detected by this single sensor. As a result, it has the least accurate position reconstruction accuracy. 
Position B lies right between two sensor pixels in the $x$ coordinate, and thus creates the optimal light distribution in this direction, leading to the best $x_1$ accuracy at the cost of compromised $y_1$ accuracy. 
For similar reasons, event position C achieves a $x_1$ accuracy between A and B but with balanced $x_1$ and $y_1$ performance. 

The predicted resolution is reported between 20 and 1,000,000 detected photons, with the lower bound corresponding to the typical number of photons detected in a single electron event in a dual-phase xenon TPC and the upper bound approximating the size of 2.5~MeV gamma ray or electron events (Sec.~\ref{sec:0vbb}). 
The resolution is observed to improve with increased photon counts in the form of $1/\sqrt{m_1}$, which is expected from Poisson distributions. 
With a 2" photosensor size, at 20 detected photons the predicted position resolution is $\sim$1~cm, which is a factor of 2 better than that demonstrated by the LUX experiment using 2" PMTs at a similar photon number~\cite{LUX2018_PositionReconstruction}. 
When the LUX LRF function is used in a similar theoretical study (Sec.~\ref{sec:lux}), the predicted SS position reconstruction resolution is within 20\% of experimental observations. 
The excellent agreement between this study and the LUX experimental result provides a strong confirmation that FI can be applied to position reconstruction in dual-phase xenon TPCs.

Figure~\ref{fig:sslimit} (right) shows the predicted $x$ position resolution for different photosensor sizes, from 3" to 6~mm in diameter. 
Here we only plot the results for event position A which has the worst sensitivity. 
As the sensor size decreases, the predicted position resolution improves approximately linearly until the sensor size becomes smaller than the light source-sensor separation. 
This is again the inverse square law dependence of the photon intensity for isotropic light emission, which dominates the light spread once the light source is further away from the photosensor plane than the sensor pixel size. 
With the TPC model described in Sec.~\ref{sec:detector}, the SS reconstruction capability is comparable for 15~mm and 6~mm sensor pixel sizes. However, a more significant improvement can be achieved for detectors that optimize the light source-sensor separation for smaller sensor pixels, to be discussed in Sec.~\ref{sec:optimize}.

\begin{figure}[!ht]
    \centering
    \includegraphics[width=0.99\textwidth]{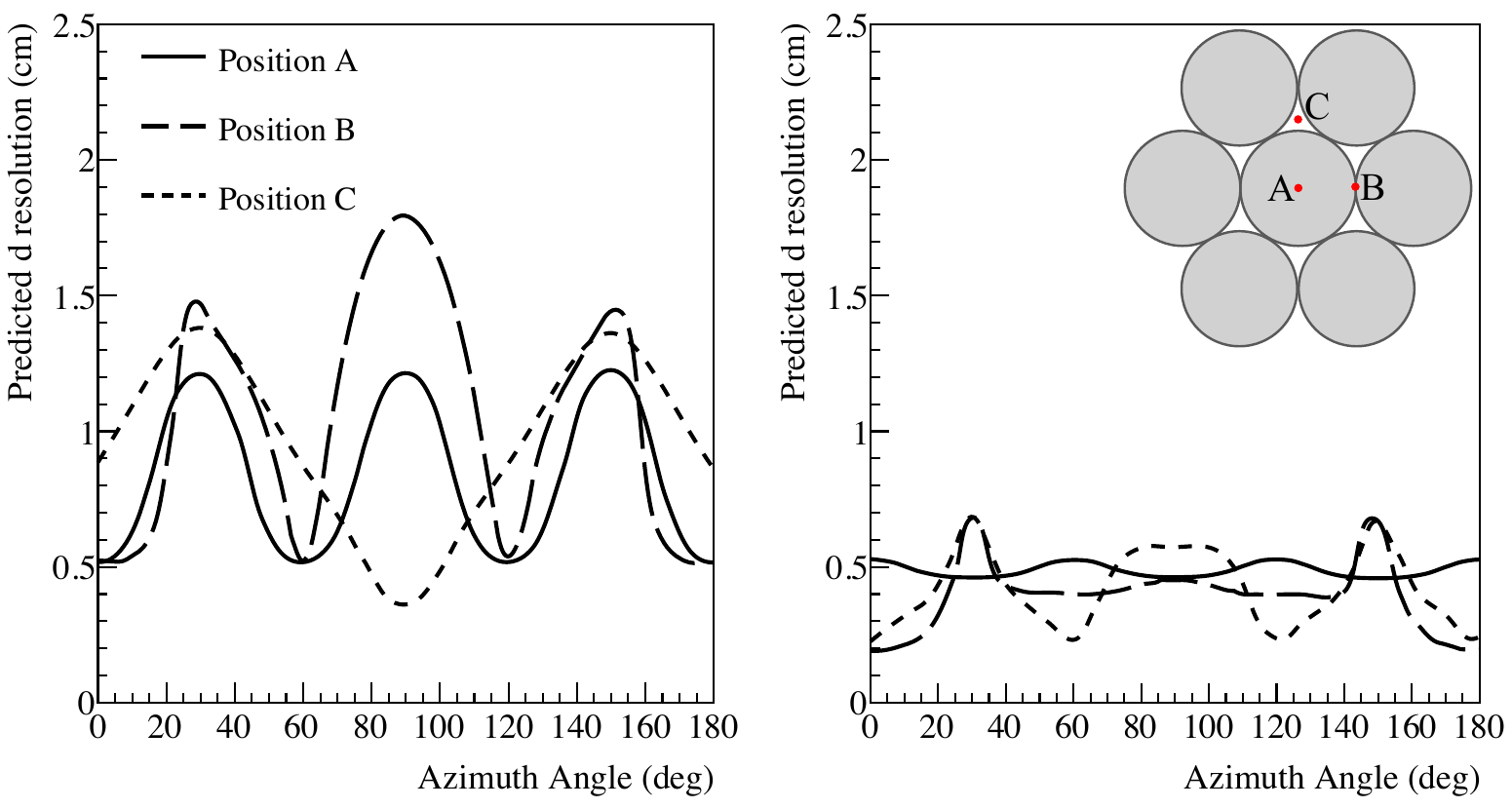}
    \caption{Predicted $d$ reconstruction accuracy with 500 total detected photons for MS2 events in our TPC detector model (2" PMTs) using FI and CRLB. \textbf{Left -} $\sigma(d)$ ($d$=1") as a function of azimuth angle of second vertex relative to the first, at event positions A (solid), B (dashed) and C (dotted); \textbf{Right-} same as left figure but with $d$=2". The total photon budget is evenly split between the two vertices. }
    \label{fig:dsang}
\end{figure}

For MS2 events, 
we use the distance between the two vertices ($d$) and the relative azimuth angle ($\phi$) to describe the location of the second vertex in the FI calculation so we can directly study the predicted accuracy in measuring $d$ that can help resolve the two vertices. 
Figure~\ref{fig:dsang} shows the theoretical limit of resolution, $\sigma(d)$, as a function of $\phi$ with $d$=1" (left) and $d$=2" (right) for 2" PMTs. Here we assume a total of 500 photons detected, shared equally between the two vertices, which corresponds to small ionization events in typical TPCs with dozens of electrons. 
The 3 curves in each figure correspond to the first event positions A (solid line), B (dashed line), and C (dotted line), respectively. 

The accuracy in reconstructing $d$ exhibits a strong dependence on $\phi$. Periodical patterns reflecting the corresponding symmetry in the photosensor layout are observed. 
Generally, when the two vertices share the same nearest sensor pixels, the $\sigma(d)$ is predicted to be large due to the partial degeneracy of the hit patterns. 
This is further supported by the reduction in $\sigma(d)$ when $d$ increases from 1" to 2", which decreases the overlap between hit patterns produced by the two vertices. 
For large $d$ separations, the dependence of $\sigma(d)$ on both $\phi$ and the first vertex position is also reduced, indicating that well-separated MS2 events are less affected by the photosensor layout. 
Similarly, the $\sigma(d)$ variations are also predicted to be small when smaller photosensors are used. 

\begin{figure}[!ht]
    \centering
    \includegraphics[width=0.99\textwidth]{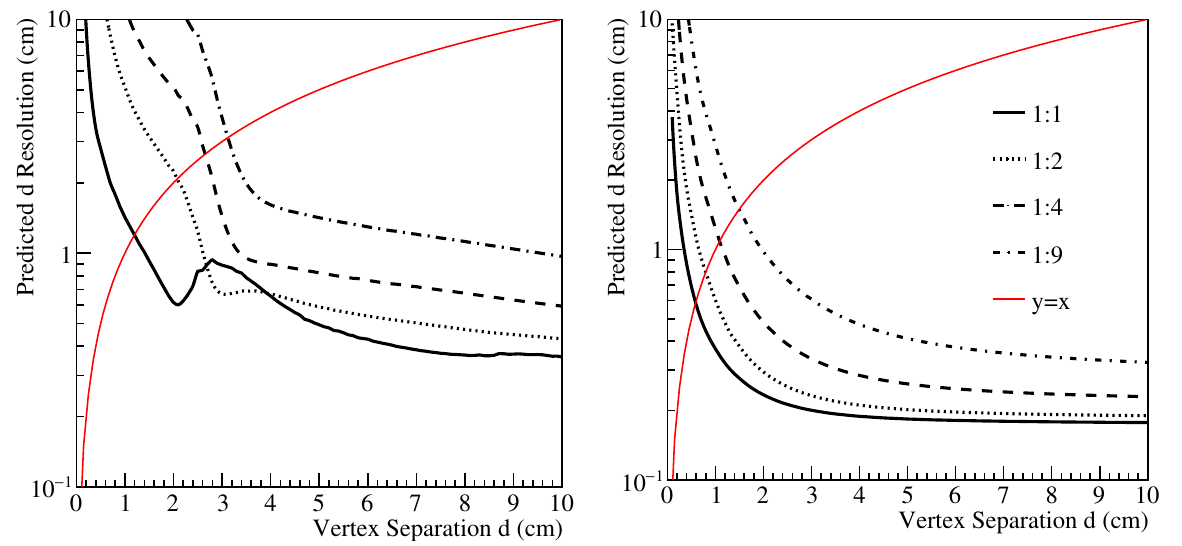}
    \caption{Predicted $\sigma(d)$ as a function of vertex separation ($d$), with 500 total detected photons for MS2 events with 2" (\textbf{left}) and 6~mm (\textbf{right}) photosensor sizes. The lines indicate different photon budget splitting between the two vertices: 1:1 splitting (solid line), 1:2 splitting (dotted), 1:4 splitting (dashed), 1:9 splitting (dot dashed). The red line represents $\sigma(d)=d$. The diverging trend near $d=0$ reflects an instability of the FI matrix and is unphysical. }
    \label{fig:dresd}
\end{figure}

To study the dependence of $\sigma(d)$ on both $d$ and the number of photons at the two vertices, we compute the average $\sigma(d)$ over all $\phi$ angles, and only choose position A for the first vertex to represent the system performance. Position A is chosen for its conservative performance and its azimuthal symmetry in the placement of the second vertex. 
Figure~\ref{fig:dresd} shows the resulting $\sigma(d)$ as a function of $d$ for 2" (left) and 6~mm sensors (right). 
We assume a total of 500 photons detected split between the two vertices with ratios of 1:1 (solid black), 1:2 (dotted black), 1:4 (dashed black), and 1:9 (dot-dashed black). 
As expected, $\sigma(d)$ improves with increasing $d$. The fine features observed in the case of 2" photosensors are a result of the second vertex crossing the boundaries between neighboring sensor pixels. 
For 6~mm studies, $\sigma(d)$ decreases monotonically with increasing $d$ without any oscillating features because the photosensor pixelation is a subdominant effect due to the large distance between the light source and the sensors.
At a fixed total photon count, large disparities in vertex brightness degrade the $d$ reconstruction, as the hit pattern produced by the dimmer vertex can be shadowed by that of the brighter one, causing its position information to be partially lost. 

For very small $d$ values, FI predicts $\sigma(d)$ values in the range of tens of cm, which are even larger than the spread of photon patterns for SS events in our TPC model. 
This prediction is unphysical and represents an instability of the FI matrix in near-degeneracy conditions. 
As discussed in Sec.~\ref{sec:fisher}, when two vertices are close to each other, a degeneracy between parameters describing the vertices emerges; changes in a parameter for one vertex may be compensated by a proportionate change in parameters for the other vertex. 
Mathematically, degeneracies of parameters cause the FI matrix, which is constructed from likelihood score covariance, to become ill-conditioned. With a complete degeneracy between two parameters, the FI matrix has eigenvalues containing 0, and is thus non-invertible; as a result, the CRLB limits would diverge. 
For these unphysical $\sigma(d)$ values, the condition number of the FI matrix, calculated as the ratio of the largest eigenvalue to the smallest, is observed to increase by orders of magnitude even with a millimeter-level change in $d$. In this work, we exclude values of $\sigma(d)$ greater than 10~cm. 
The validity of this approach will be further studied with the simulation and reconstruction efforts presented in Sec.~\ref{sec:methods}. 

We define a benchmark MS2 resolution $\delta_{d}$ as the value $d$ at which $\sigma(d)=d$, i.e., when the precision to reconstruct the distance between the two vertices is equal to the distance itself. 
At this distance, there is a $\sim$16\% probability for an unbiased estimator to reconstruct $d$ as negative~\footnote{Our FI calculation does not consider bounds for parameters (unbiased), so the inter-vertex distance $d$ (and the photon count $m$) may take on both positive and negative values. } if we assume a Gaussian distribution for the obtained $d$ value. The same estimator would also predict a mean $d$ value of 0 for SS events. 
Therefore, a cut of $d<0$ would accept 50\% SS events while allowing for a 16\% MS leakage, which will be used as the benchmark for comparison with reconstruction results in Sec.~\ref{sec:methods}. 
In Fig.~\ref{fig:dresd}, the benchmark resolution values are obtained at the intercepts of the $y=x$ line with the $\sigma(d)-d$ curves. 
For 2"-size sensors $\delta_{d}$=1.2~cm when 500 photons are detected and the photons are equally distributed between the two vertices. 
$\delta_{d}$ increases to 3.1~cm when one vertex is $\sim$10\% as bright as the other, even when the same number of total photons is detected. 
Similar observations are made for the 6~mm sensor pixel sizes, except that both $\sigma(d)$ and its dependence on sensor layout is reduced thanks to the finer pixelation of the photosensor array. 

\begin{figure}[!ht]
    \centering
    \includegraphics[width=0.99\textwidth]{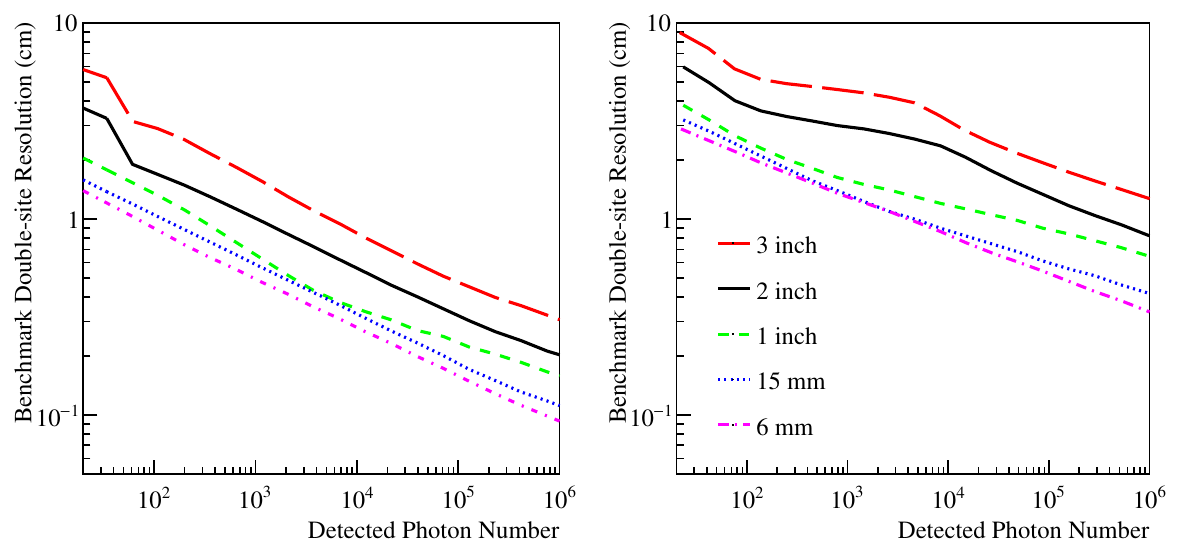}
    \caption{Benchmark $d$ reconstruction resolution as a function of number of photons detected for 1:1 (\textbf{left}) and 1:9 (\textbf{right}) photon budget splitting between the two vertices. Different lines indicate different photosensor sizes - from top to bottom: 3", 2", 1", 15~mm, and 6~mm circular photosensors. }
    \label{fig:dresn}
\end{figure}

Figure~\ref{fig:dresn} shows the  benchmark MS2 resolution $\delta_{d}$ calculated as a function of the number of photons detected, for different photosensor pixel sizes: 3", 2", 1", 15~mm, and 6~mm (from top to bottom); the figure on the left assumes a 1:1 photon splitting between the two vertices while the figure on the right assumes a 1:9 photon splitting. 
Generally, collecting more photons and using smaller sensor pixels reduce uncertainties in the estimated separation between the two vertices, but the improvement diminishes when the sensor size becomes smaller than the distance between the light source and the photosensor array. This is consistent with the results for the SS event reconstruction predictions. 
In addition, vertices with comparable brightness are easier to resolve than those with large disparities; a deterioration factor of 3 is observed for the 1:9 photon splitting scenario compared to the equal photon splitting scenario. 

Lastly, we point out that the smallest position separation between two vertices that can be resolved is significantly larger than the predicted position resolution for SS events, when all considered vertices have the same photon number. For example, in Fig.~\ref{fig:sslimit}, with 250 photons detected, a SS event can be reconstructed with 3~mm accuracy for 2" photosensors, but when two vertices, each with detecting 250 photons, are present, the benchmark resolution of the separation between the two is 1.2~cm. 
This factor of 4 discrepancy is much larger than that of $\sqrt{2}$ expected by subtracting one vertex position from the other. 
This results from the ambiguity in associating a detected photon with a certain vertex. When there is only one vertex in the system, each detected photon carries definitive information about its origin, whereas for multiple vertices hit patterns significantly overlap and photons from both vertices may have similar probability to hit the same sensor pixel. 
This contributes to the uncertainty in locating the position for all overlapping vertices.

\begin{figure}[!ht]
    \centering
    \includegraphics[width=0.65\textwidth]{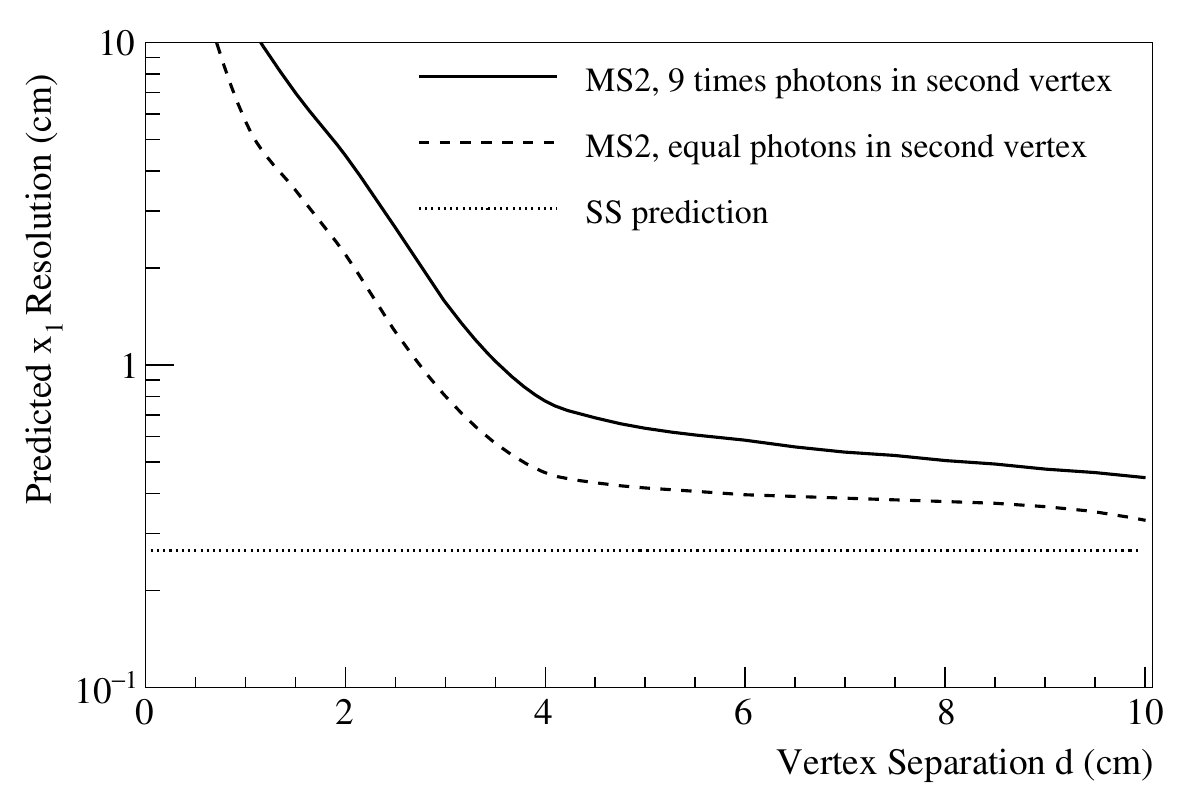}
    \caption{Predicted $x_1$ reconstruction resolution for a vertex of 250 detected photons as a function of distance to a second vertex of equal (dashed line) and 9 times (solid) photon strength, in comparison with the SS position resolution (dotted); here we assume a 2" photosensor size and the first vertex locates right below the center of a sensor pixel (Position A), while the second vertex moves along the $x$ axis. The diverging trend near $d=0$ reflects an instability of the FI matrix and is unphysical. }
    \label{fig:xresvsd}
\end{figure}

To clearly demonstrate this point, we study the position reconstruction uncertainty for a vertex with and without a second vertex nearby. Fig.~\ref{fig:xresvsd} shows the theoretically predicted position resolution for a vertex with 250 detected photons as a function of the distance at which a second vertex of equal (dashed line) and 9 times (solid line) photon count is placed. 
When the two vertices are far apart, the predicted position resolution approaches the SS prediction (dotted line), but when the two vertices are close to each other, the best reconstruction accuracy deviates substantially from the SS limit. 

\section{Reconstruction and validation}
\label{sec:methods}

In this section, we test position reconstruction and SS/MS discrimination with our simple TPC model. Results obtained from methods such as centroiding, principal component analysis (PCA), maximum likelihood estimator (MLE), and convolutional neural network (CNN) are compared to theoretical predictions in Sec.~\ref{sec:sensitivity} and available experimental results. 
All reconstruction studies are performed with 2" PMTs in the TPC model, and for MS events we only consider MS2 events with equal photon split between the two vertices. 
However, the same methods can be applied to other detector models studied in Sec.~\ref{sec:theory}, or more realistic detectors provided their light responses are known from simulation or direct measurements. 

\subsection{Single-site position reconstruction}
\label{sec:ssmethods}

\begin{figure}[!ht]
    \centering
    \includegraphics[width=0.47\textwidth]{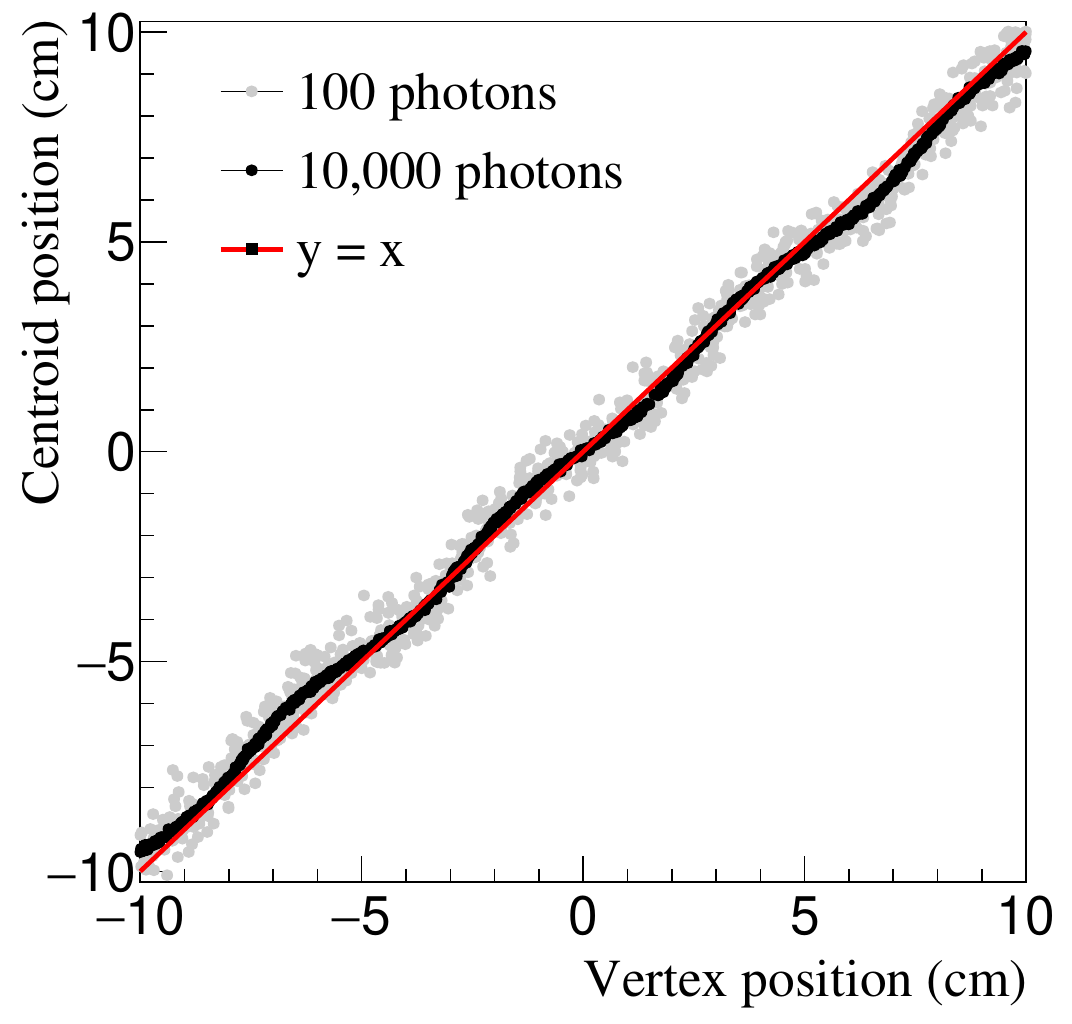}
    \includegraphics[width=0.51\textwidth]{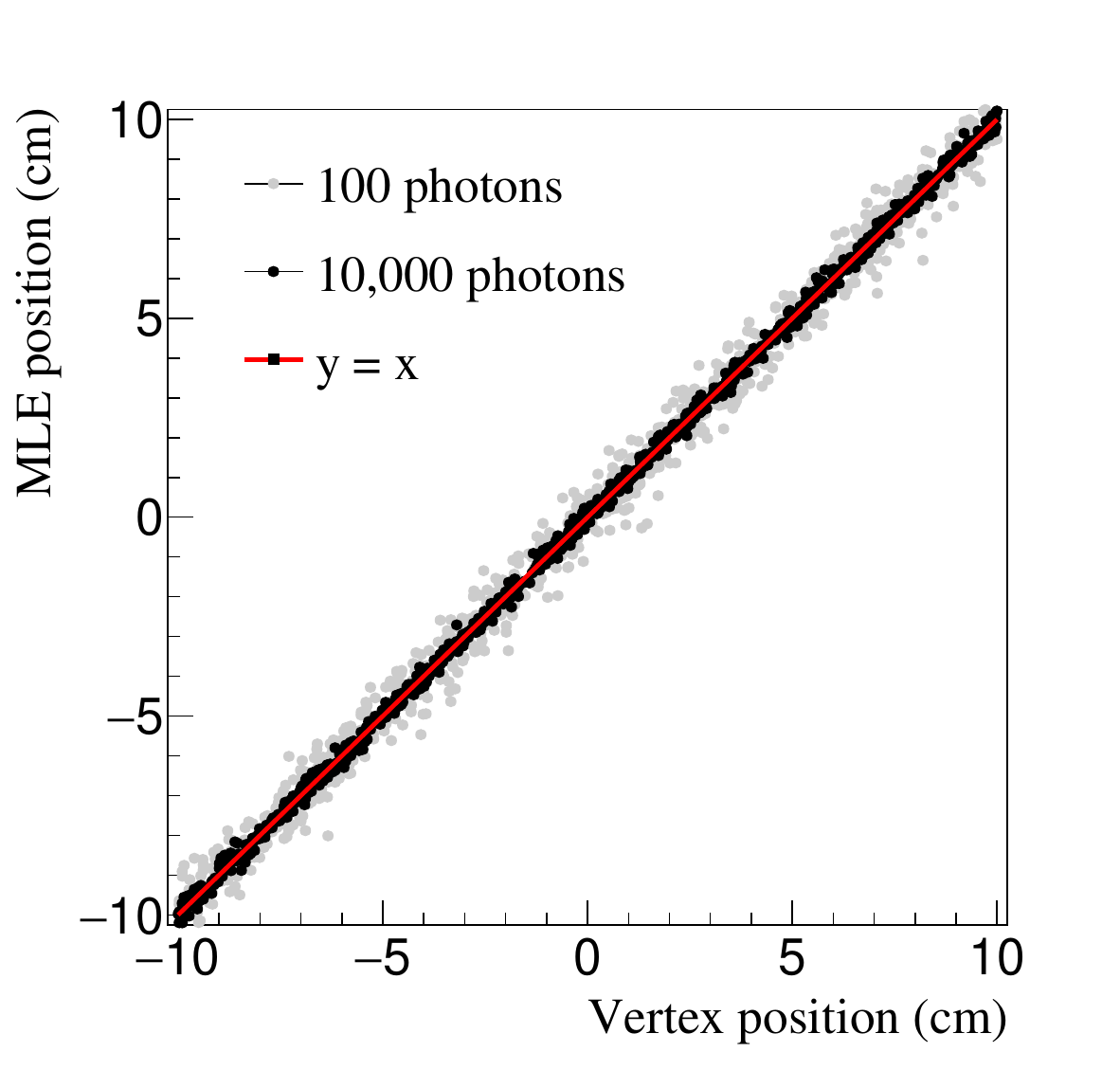}
    \caption{\textbf{Left} - centroid $x_1$ position reconstructed for SS events ($y_1$=0) with 100 (gray) and 10,000 (black) photons detected in the TPC model using 2" sensors; \textbf{Right} - the same as left but with the maximum likelihood reconstruction. The coordinates are defined in Fig.~\ref{fig:setup}. }
    \label{fig:centroidmle}
\end{figure}

For a given sensor hit pattern, the mean position of all recorded photon hits provides a straightforward estimate of the event location. This method is often referred to as centroiding or center of mass. 
Fig.~\ref{fig:centroidmle} (left) shows the $x_1$ centroid position for SS events as a function of the true $x_1$ position for 100 (grey points) and 10,000 (black points) detected photons, using a 2" sensor pixel size. True $x_1$ positions are randomly sampled between -10~cm and 10~cm while keeping $y_1$ at 0. The sensor hit patterns are then generated using a toy Monte Carlo simulation that samples the parameterized LRF. 
Generally, the obtained centroid position closely tracks the true event position, and the precision improves with increasing number of photons detected. 
However, even with 10,000 detected photons, the centroid position may still deviate from the true event position with a pattern corresponding to the sensor layout. 
This discrepancy originates from the inaccurate photon detection position estimated using discrete pixel center positions and can be partially mitigated by reducing the pixel size or increasing the light source-sensor separation. For actual detectors with a finite size of sensor array, the centroid position will carry an additional bias from the asymmetric detection of photons, which usually pulls the reconstructed event position toward the center of the detector. 

The aforementioned bias with centroiding estimation is addressed in more sophisticated methods such as MLE. 
As demonstrated by the ZEPLIN-III and LUX experiments~\cite{posReconstruct_Solovov2012,LUX2018_PositionReconstruction}, MLE can accurately reconstruct the position of SS events  using a sensor's response to light sources of known locations. 
Given a sensor response function of $\eta(r)$, the likelihood function for a given measurement can be calculated as
\begin{equation}
    \mathcal{L} (X |\Theta ) = \prod_{k}^K \frac{\eta_k^{n_k} e^{-\eta_k}}{n_k!}
\end{equation}
where $X$ represents all measured data including the positions of the photosensor pixels and the detected photon count in each pixel ($n_k$  for sensor $k$), $\Theta$ represents the parameters of interest including the brightness and $x-y$ positions of event vertices, and $\eta_k(r)$ is the Poissonian mean of the expected photon count in the $k$th pixel for given $\Theta$ values. 
As discussed in Sec.~\ref{sec:detector}, we use an analytical LRF that only depends on the horizontal distance between the light source and the sensor pixel center, but the method can be easily generalized for detectors with more complex responses. 
In practice, instead of maximizing the likelihood, the negative log-likelihood -log$\mathcal{L}(X | \Theta)$ is minimized to find the best estimate of parameters. 

\begin{figure}[!h]
    \centering
    \includegraphics[width=0.56\textwidth]{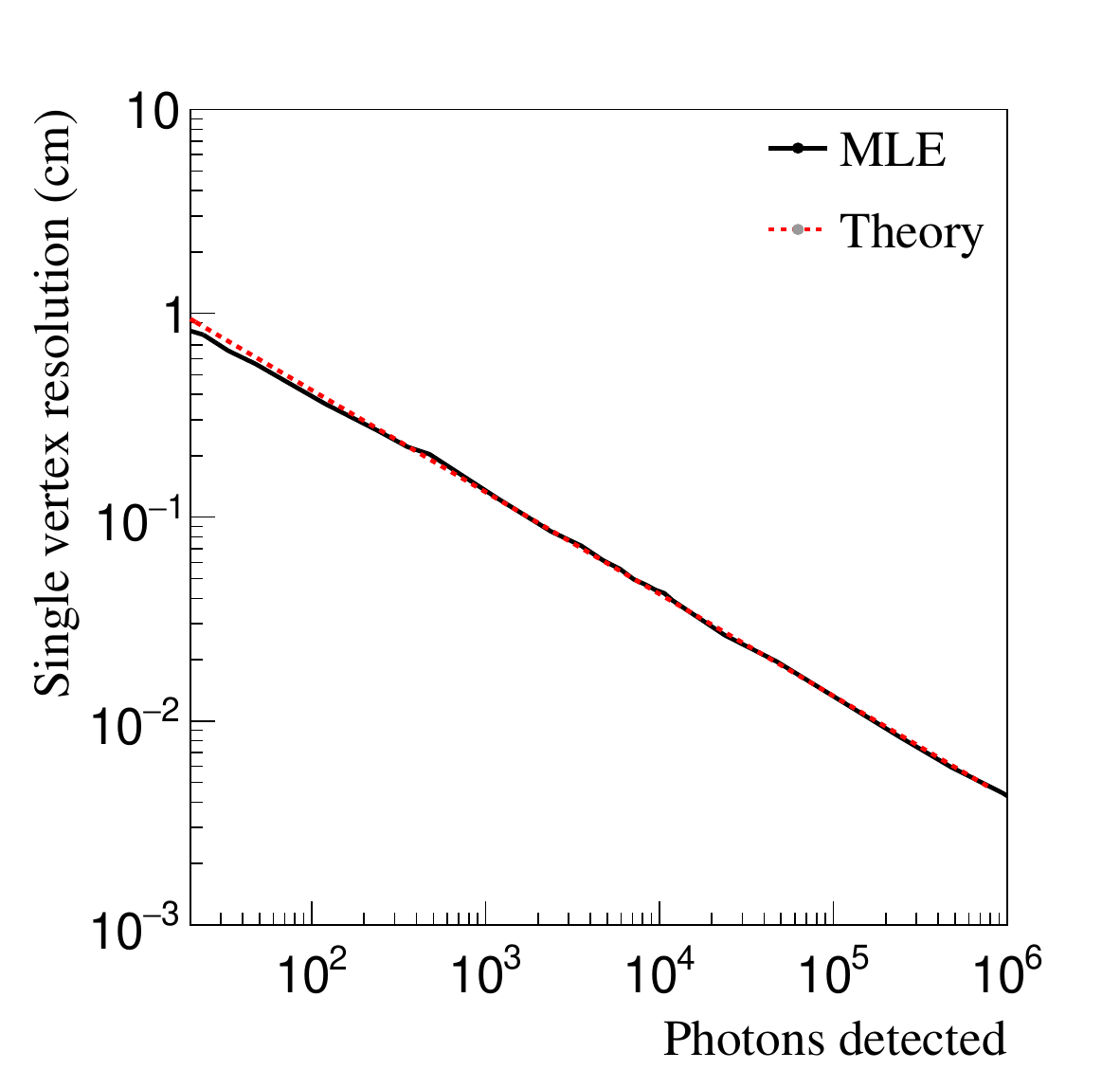}
    \caption{Reconstructed position resolution for SS events right under a PMT (position A) in the TPC model with 2" PMTs as a function of photons detected. }
    \label{fig:mle}
\end{figure}

Figure~\ref{fig:centroidmle} (right) shows the MLE-reconstructed $x_1$ position for SS events of 100 and 10,000 detected photons for 2" photosensors. 
Compared to the centroiding result in the left subfigure, the MLE reconstruction uncertainty is significantly reduced, and more importantly, the reconstruction does not exhibit any significant deviation from the true event position. 
This is more evident for the result obtained with 10,000 detected photons which has much reduced statistical fluctuation. 
Figure~\ref{fig:mle} shows the resolution of MLE reconstruction positions as a function of detected photons (solid black), in comparison to that predicted by FI (dashed red),  
the two of which are nearly identical. Because both FI and CRLB are based on likelihood and the MLE reconstruction maximizes the likelihood, it is not surprising that the theoretical limit can be achieved for a relatively simple SS system when there is no additional uncertainty. 
In an actual detector where the LRF is not exactly known or its system dependence is not fully captured by the model, the MLE reconstruction accuracy will be worse than the theoretical limit.

\subsubsection{Validation with LUX experimental results}
\label{sec:lux}

The LUX experiment operated a dual-phase xenon TPC to search for dark matter interactions between 2012 and 2016~\cite{LUX_2016DMresults}. The LUX TPC used 61 2" PMTs in the xenon gas phase and a mirrored array in the liquid to collect xenon scintillation and electroluminescence light. In this configuration, LUX reported a $x-y$ position resolution of $\sim$2~cm for single electron pulses with $\sim$20 detected photons using the MLE method~\cite{LUX2018_PositionReconstruction}. 

\begin{figure}[!ht]
    \centering
    \includegraphics[width=0.49\textwidth]{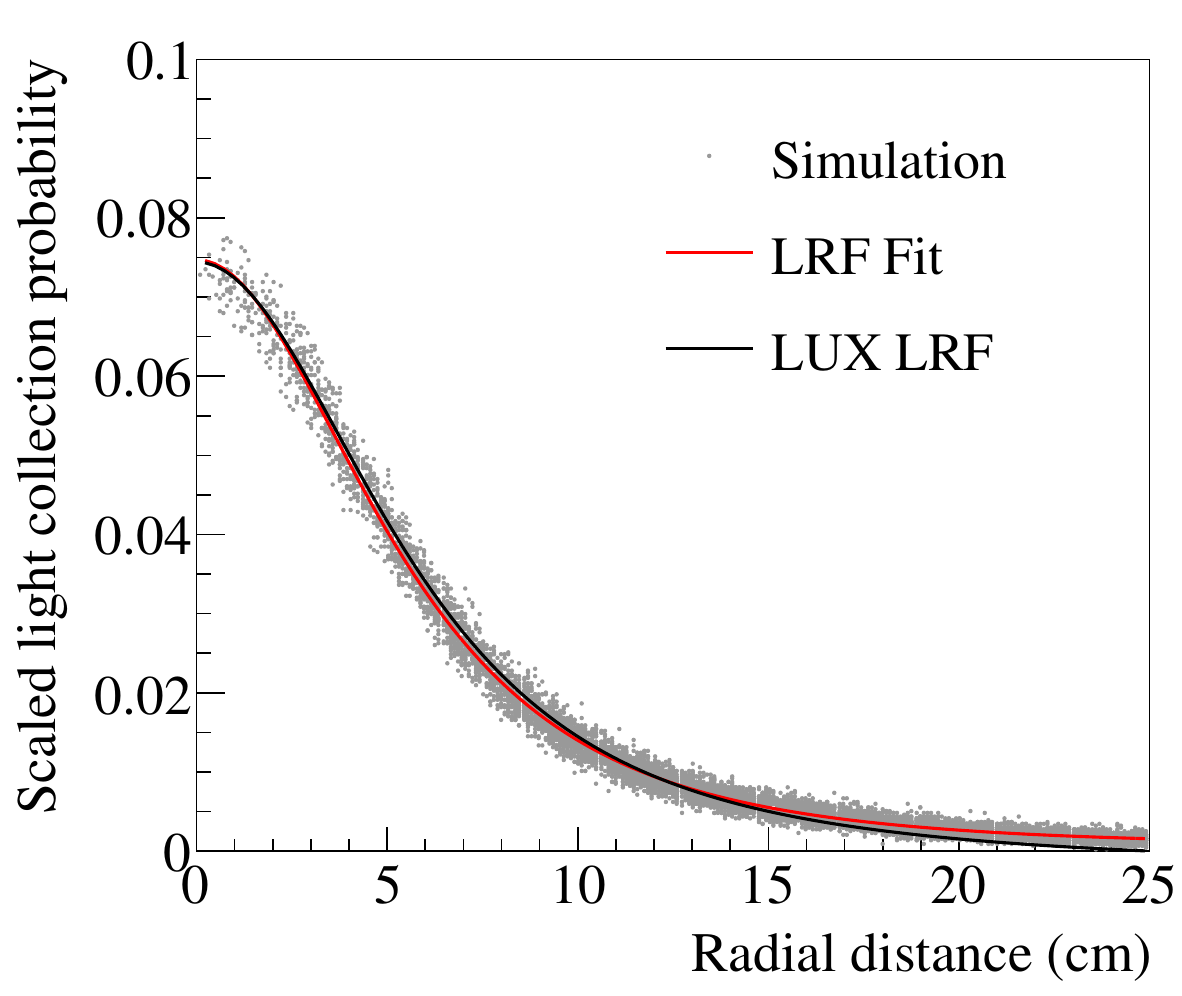}
    \includegraphics[width=0.5\textwidth]{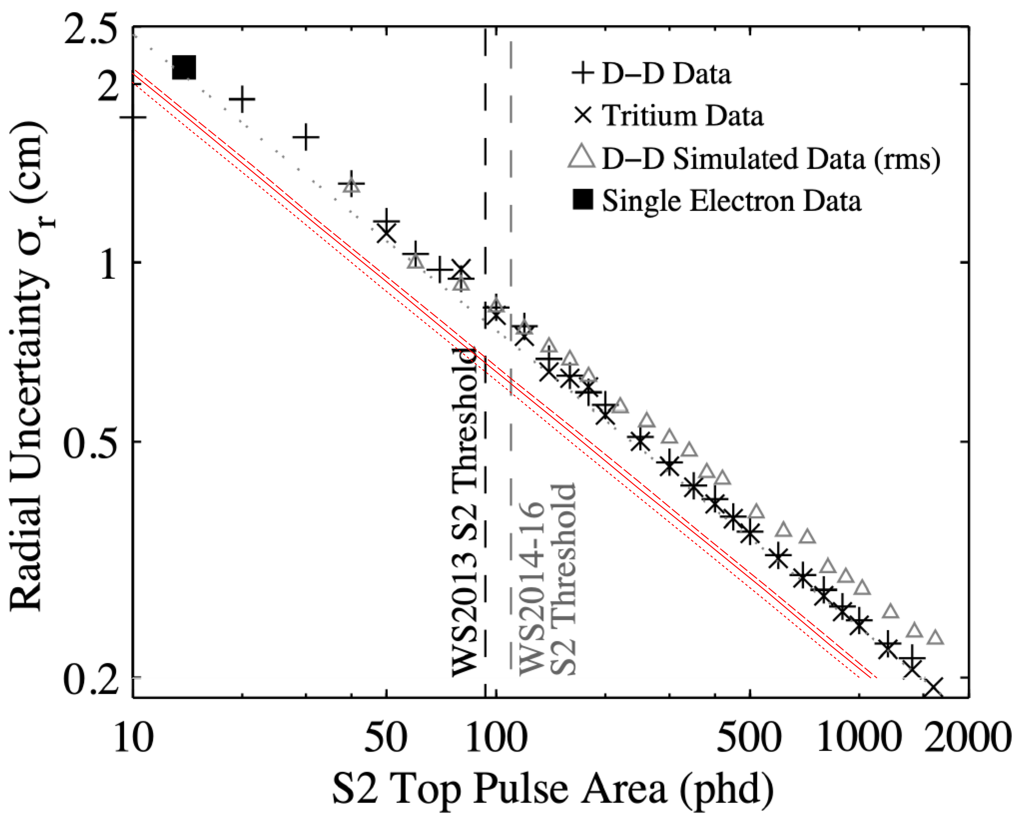}
    \caption{\textbf{Left} - simulated light collection efficiency in this work in comparison with the radial component of the LUX LRF~\cite{LUX2018_PositionReconstruction}; the best fit to our simulation using the LUX LRF formula is also shown. \textbf{Right} - predicted SS event reconstruction resolutions as a function of detected photons for 3 event positions (see text), plotted on top of experimental result from the LUX experiment. The LUX parameters are used in our simulation and sensitivity predictions. }
    \label{fig:luxcomp}
\end{figure}

The LUX TPC was 50~cm in diameter with a 0.5~cm electroluminescence region between the liquid surface and the TPC anode, and the PMT array was placed 6~cm above the anode. 
Fig.~\ref{fig:luxcomp} (left) compares the radial component of the LUX LRF~\cite{LUX2018_PositionReconstruction} with our simulated LRF using the LUX TPC configuration~\footnote{Therefore, this TPC model is different from our default 2" PMT TPC model used in other validations.}. 
In addition to adjusting the TPC dimensions, we also increase the PMT center-to-center separation to 6.125~cm from the nominal 2" diameter to be consistent with LUX where the extra space was needed to support the PMTs. 
Despite our simplified optical model, the two LRF functions are nearly identical, which provides a strong validation of our optical simulation model.  

Using the LUX radial LRF provided in \cite{LUX2018_PositionReconstruction}, we carried out a similar theoretical study to predict the SS event reconstruction resolution in the LUX detector using FI. 
Here we only consider SS events near the center of the detector so the polar component of the LUX LRF (not included in this study) is small and can be neglected. 
Fig.~\ref{fig:luxcomp} (right) shows the predicted SS reconstruction resolution curves for events right below the center PMT (red solid line), between 2 center PMTs (red dashed line) and at the center of 3 PMTs (red dotted line), which are referred to as positions A, B, and C, respectively in Fig.~\ref{fig:sslimit}, on top of experimental results reported by LUX. 

Because of the large light source-PMT array distance in the LUX configuration, the detector's position resolution is not sensitive to the event position relative to the PMT layout, and thus the predictions for the 3 representative event positions are very similar. 
Notably, the FI predictions are only $\sim$20\% better than those achieved by the LUX MLE reconstruction for actual data, strongly supporting the FI results as practical upper limits.  
The difference between FI predictions and LUX results is attributed to detector effects such as light reflection off detector walls, non-unit and non-uniform PMT quantum efficiencies, and additional PMT noise to Poissonian fluctuations.  
These experimental factors are not included in our theoretical studies. As a result, the simulation-based validations, which also ignore these effects, may produce better agreement with theory than for real experiments. However, the close match between simplified FI predictions and the LUX experimental results confirms that the theoretical precisions can be reasonably approached in real experiments. 

\subsection{Multi-site background rejection in the transverse plane}
\label{sec:msmethods}

In this section, we seek to validate FI predictions for MS2 events using the PCA, MLE and CNN methods. Instead of estimating the inter-vertex separation for such events, we focus on the power of these methods in differentiating SS and MS2 events, which is the main goal of many physics applications.  
Similar to the SS reconstruction study, we simulate MS2 events by randomly drawing photons from two vertices of assigned $x-y$ positions and estimating the PMT occupancy using the parameterized LRF. 
The PMT size is assumed to be 2" and the total number of photons are evenly split between the two vertices. 

\subsubsection{Principal component analysis}
\label{sec:pca}

MS2 events are expected to produce a photon hit pattern with larger variance along the direction connecting the two vertices (hereafter referred to as the principal direction) compared to its orthogonal direction. We use the Principal Component Analysis (PCA) to identify the principal axis direction and to quantify this asymmetric variance for SS and MS events. Mathematically, PCA diagonalizes the covariance matrix and rotates the coordinates to align with the principal direction. In addition to computing the rotation angle, PCA also calculates the eigenvalues of the covariance matrix, which correspond to the variance in photon spread in the principal and secondary directions. 
We will refer to the square roots of these variance values as the principal and secondary spreads. 

\begin{figure}[!ht]
    \centering
    \includegraphics[width=0.49\textwidth]{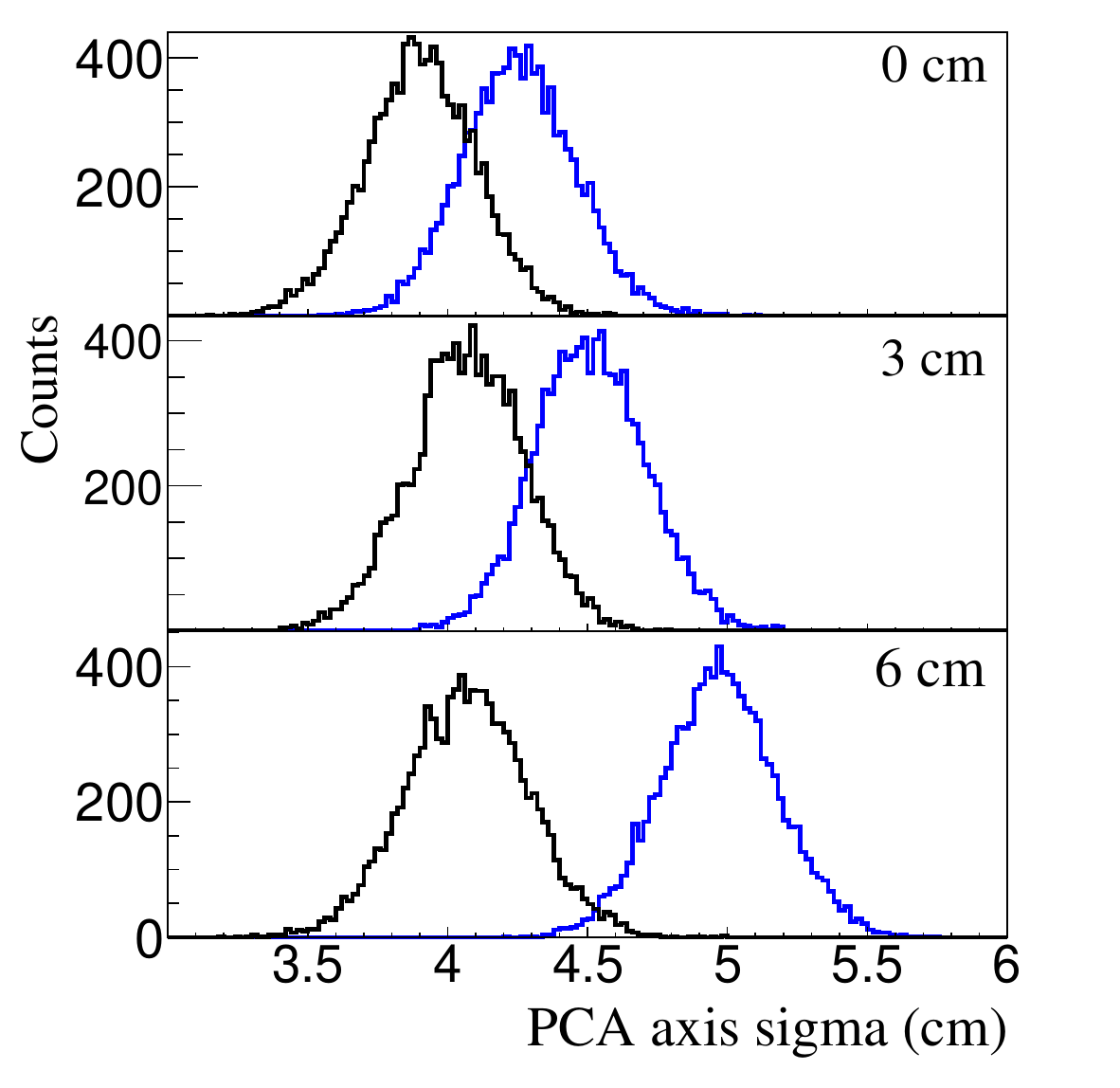}
    \includegraphics[width=0.49\textwidth]{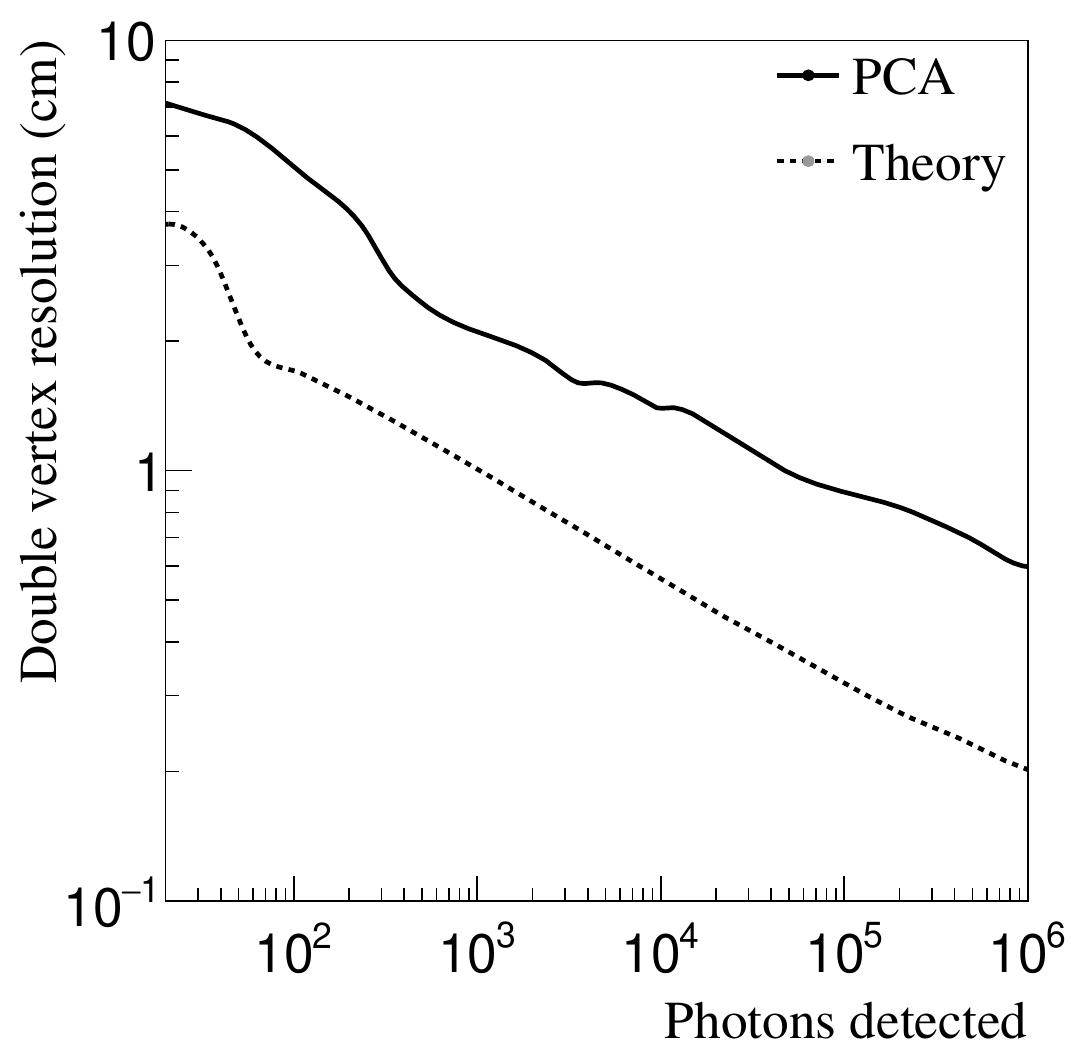}
    \caption{\textbf{Left} - the spread of photon hit along the principal (blue) and secondary (black) axis for SS (top) and MS2 events with 3~cm (middle) and 6~cm (bottom) separation, with a total of 500 photons detected. \textbf{Right} - the MS2 reconstruction resolution (solid) with PCA for different numbers of detected photons overlaid on the benchmark resolution calculated from FI (dash). In both figures, we assume a 2" photosensor pixel size and an even-splitting of photon count between the two vertices. }
    \label{fig:pca}
\end{figure}

Figure~\ref{fig:pca} (left) shows the distributions of principal (blue) and secondary (black) spread for double vertices separated by 0, 3~cm and 6~cm with a total of 500 detected photons. Here we assume a sensor pixel size of 2" and an 1:1 photon splitting between the two vertices. 
Due to statistical fluctuations, the PCA algorithm proposes principal and secondary directions even for SS events (0 separation between vertices). 
Therefore, the appearance of asymmetric variance, by itself, cannot be used to discriminate SS and MS events. 
Instead, we use the displacement of the mean principal spread from the SS mean to estimate how well MS events may be distinguished from SS events. 
This method takes statistical fluctuations of SS variance distribution into consideration. When this displacement equals the width of the principal spread distribution, 
we obtain a 50\% SS acceptance while allowing 16\% MS contamination. 
The corresponding vertex separation can be directly compared to the benchmark MS2 resolution $\delta_{d}$ defined in Sec.~\ref{sec:sensitivity}. 
The PCA-derived $\delta_{d}$ values are compared to that predicted by FI, as a function of detected photon count, in Figure~\ref{fig:pca} (right). 
The PCA-derived resolution values follow the same trend as the theoretical predictions, but are approximately twice larger at the same photon statistics. 

\subsubsection{Likelihood ratio analysis}
\label{sec:lrf}

MLE can reconstruct the inter-vertex separation of MS2 events. 
However, when the total number of vertices is unknown as in a real MS event, the MLE optimization problem can become ill-defined. Therefore, 
we only quantify the goodness of MLE fits for simulated MS2 events with a likelihood ratio, which is defined as
\begin{equation}
    q = - 2 \textrm{log} \left(\frac{\mathcal{L}(\hat m,\hat x_1,\hat y_1,f_2=0 | n_k )}{\mathcal{L}(\hat m,\hat x_1,\hat y_1,\hat f_2,\hat d,\hat \phi | n_k) } \right)
    \label{eq:llr}
\end{equation}
where the ``hat'' symbol indicates the best-fit parameter values. 
$q$ measures the improvement of the fit quality with a MS2 hypothesis over a SS hypothesis. 
It is also a statistic that is invariant under change of variables, and can indicate MLE's capability in resolving MS2 events. 
More importantly, despite its construction, this method retains some power for SS/MS event discrimination even when the number of vertices in an event is not exactly 2, as demonstrated in Sec.~\ref{sec:0vbb}. 

\begin{figure}[!ht]
    \centering
    \includegraphics[width=0.49\textwidth]{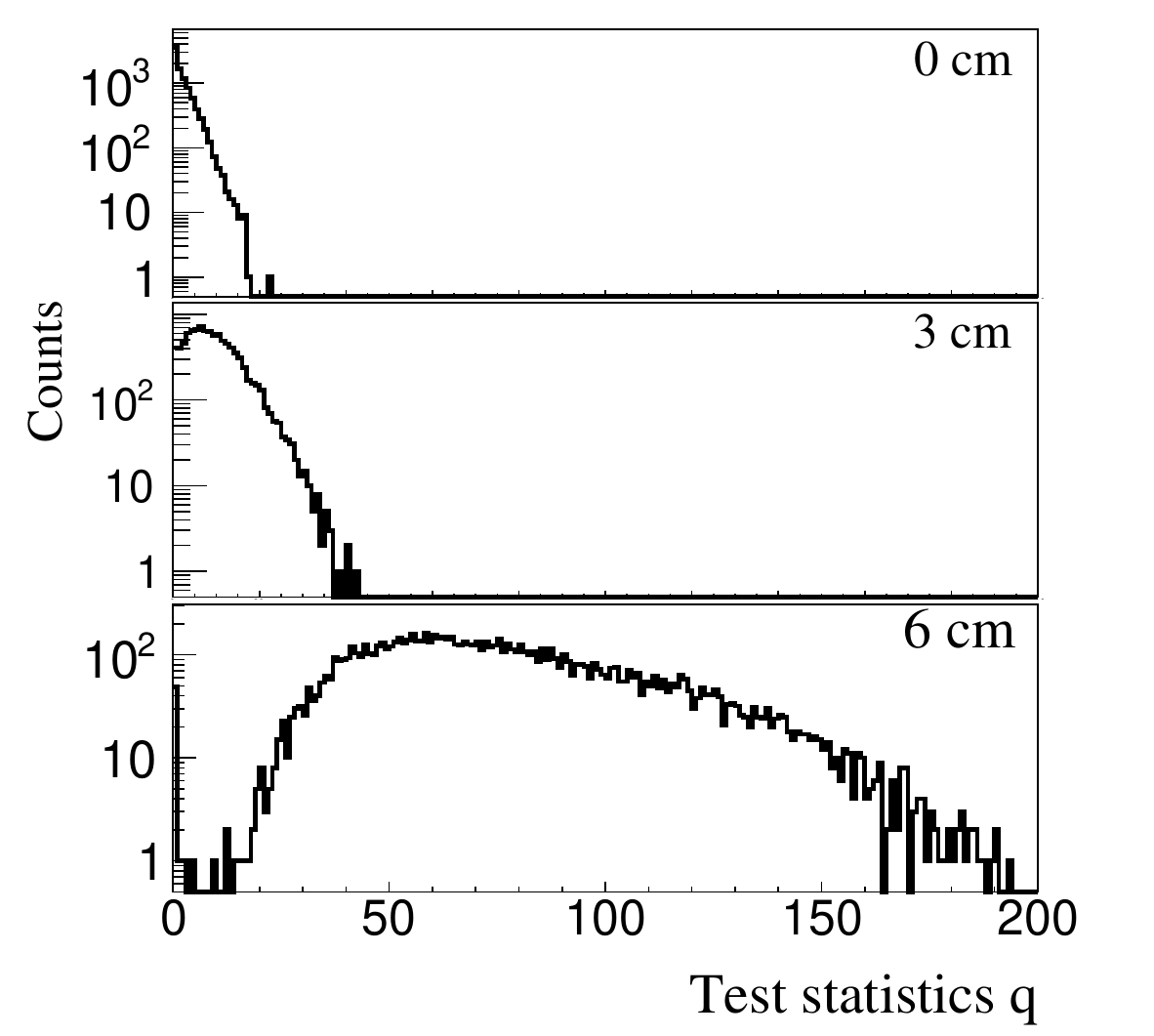}
    \includegraphics[width=0.49\textwidth]{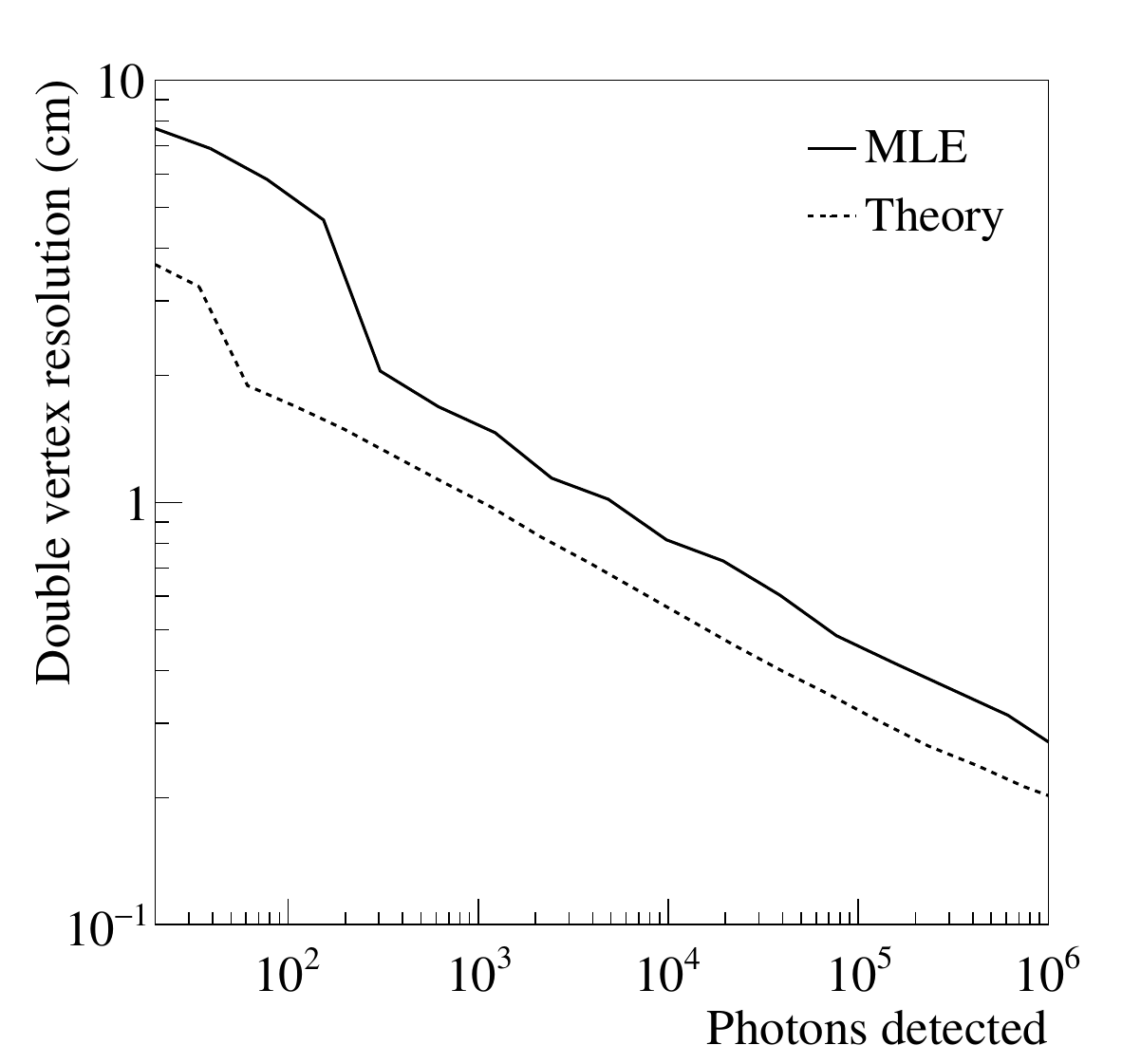}
    \caption{\textbf{Left} -  distributions of MLE test statistic $q$ for SS (top) and MS2 events with 3~cm (middle) and 6~cm (bottom) separation. Here we assume an even-splitting of 500 photons detected by 2" photosensors for two vertices. \textbf{Right} - MLE-based MS2 resolution (solid line, defined in text) as a function of the number of photons detected, in comparison with the theoretical prediction (dotted line, from Fig.~\ref{fig:dresn}).}
    \label{fig:mledisc}
\end{figure}

Because the MS2 fit includes additional parameters and should, on average, obtain a larger likelihood value, the likelihood ratio should be bounded between 0 and 1. Therefore, $q$ should be a positive number and the factor of 2 is added so the distribution of $q$ for SS events follow a $\chi^2$ distribution (Wilk's theorem). 
The $q$ distribution for SS events, and MS2 events with inter-vertex distances of 3 and 6~cm are shown in Fig.~\ref{fig:mledisc} (left). 
Here we conveniently use the PCA-estimated principal direction and variance to initialize the MLE fit.
For SS events, the two likelihood values should be close to each other, yielding $q\sim$0, while for resolvable MS2 events the MS2 fit should return a much larger likelihood, causing $q>0$. 
The MLE's ability to distinguish MS and SS events resides in the deviation of the $q$ distribution from 0 for MS2 events. 

To obtain a characteristic MS2 resolution that can be compared to the benchmark theoretical prediction, we choose the inter-vertex separation value $d$ that leads to 16\% MS2 contamination when a cut on $q$ value accepts 50\% of SS events. 
For 500 detected photons the characteristic MS2 resolution is estimated to be 1.7~cm, which is slightly larger than the theoretical limit of 1.2~cm. 
The characteristic resolution dependence on the number of detected photons (assuming 1:1 splitting between two vertices) is shown in Fig.~\ref{fig:mledisc} (right), in comparison with that predicted by FI and CRLB (Fig.~\ref{fig:dresn}). 
The MLE is observed to outperform the PCA for most of the scenarios tested. 

\subsubsection{Convolutional neural network}
\label{sec:cnn}

Machine learning (ML) techniques are commonly used in classification problems, and have been applied to signal/background discrimination in both single-phase \cite{DNN_EXO200_2018} and dual-phase~\cite{LUX_Migdal_2021,ML_pulse_classifier_2022} TPCs. These techniques enable subtle features of event topology, beyond 
traditional analysis parameters describing event energies and positions, to be possibly identified. 
Convolutional neural nets (CNNs), in particular, have found use due to their ability to leverage spatial correlations between features in image-like data by adding convolutional layers to fully connected neural nets. 
In this section, we demonstrate the use of a CNN for SS/MS discrimination in our TPC model. 

When simulating events in a large TPC instrumented with 2" PMTs, 
SS event positions are drawn uniformly from within a central disk. For MS2 events, the distance from the first vertex to the second, and the angle defining the direction of the second vertex are both sampled from uniform distributions. While separation between vertices in physical events, such as those produced by Compton scatters of gamma rays, are often exponentially distributed, we use the uniform distribution to ensure training data span a sufficient range of separations. 
As done in studies presented early in this section, the mean number of detected photons at each sensor is sampled from a Poisson distribution with a mean determined from an analytical LRF and the radial distance to the event vertex. 

\begin{figure}[!t]
    \centering
    \includegraphics[width=\textwidth]{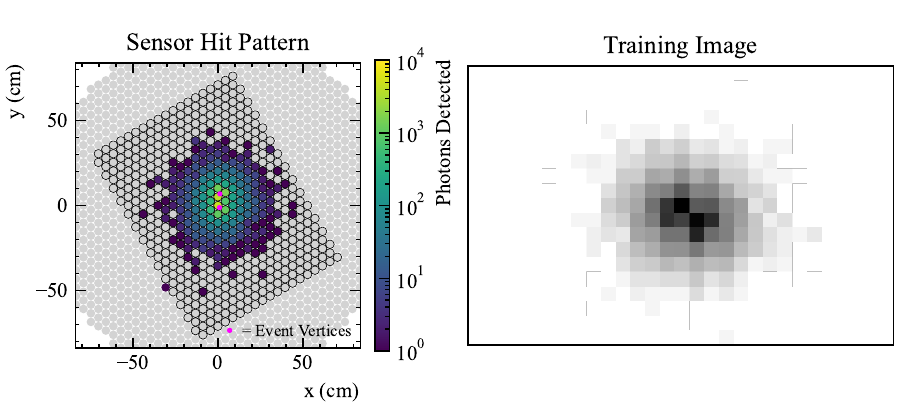}
    \caption{\textbf{Left} - a sample sensor hit pattern for a double scatter near the center of the detector. The magenta dots indicate the positions of the two vertices while the colorbar shows the number of photons detected on each sensor. Sensors with a black outline are included in the resulting image. \textbf{Right} - the gray-scale image passed to the CNN corresponding to the hit pattern shown on the left.}
    \label{fig:cnn_image_prep}
\end{figure}

The sensor hit pattern in each simulated event produces an image with which we can train or test the CNN. 
First, the sensor nearest the photon centroid is chosen as the center of the image. Then, a rectangular image is constructed by taking a fixed number of rows and columns along a selected direction. 
Due to the six-fold rotational symmetry of the sensor array arising from the hexagonal packing of the circular sensors, we choose the direction in which the second moment of the light detected is maximal. 
This ensures that the variance in light collection is predominantly along the horizontal axis of each image, reducing the number of patterns the CNN must learn.
The resulting rows of sensors are then mapped to a rectangular grid of pixels by interpolating the pixel value in the alternate columns. To ensure that the CNN does not use the total energy of an event, and leverages only the spatial information, the number of detected counts at each pixel is then normalized by the sum over the full image. 
This process is illustrated in Fig. \ref{fig:cnn_image_prep}, which shows a sample sensor hit pattern on the left and the resulting image used to train the CNN on the right.

The obtained 2D images are then processed using a ResNet-14 \cite{ResNet_2016} architecture with 8, 16, and 32 filters at each convolutional layer. 
While the ResNet architecture can be scaled to larger training datasets for more complex tasks, a simpler CNN architecture such as a small-scale VGG \cite{VGG_2014} was found to yield a similar performance in our study.
The CNN has $\sim\!4.5\times10^4$ parameters, and the training data consists of $10^5$ images each for the SS and MS categories. 
The choices ensure at least as many training samples as trainable parameters. The CNN is trained using a binary cross-entropy loss function to categorize events as SS or MS. Learning rate scheduling and early stopping were implemented to maximize performance of the classifier on the validation data.

\begin{figure}[!t]
    \centering
    \includegraphics[width=\textwidth]{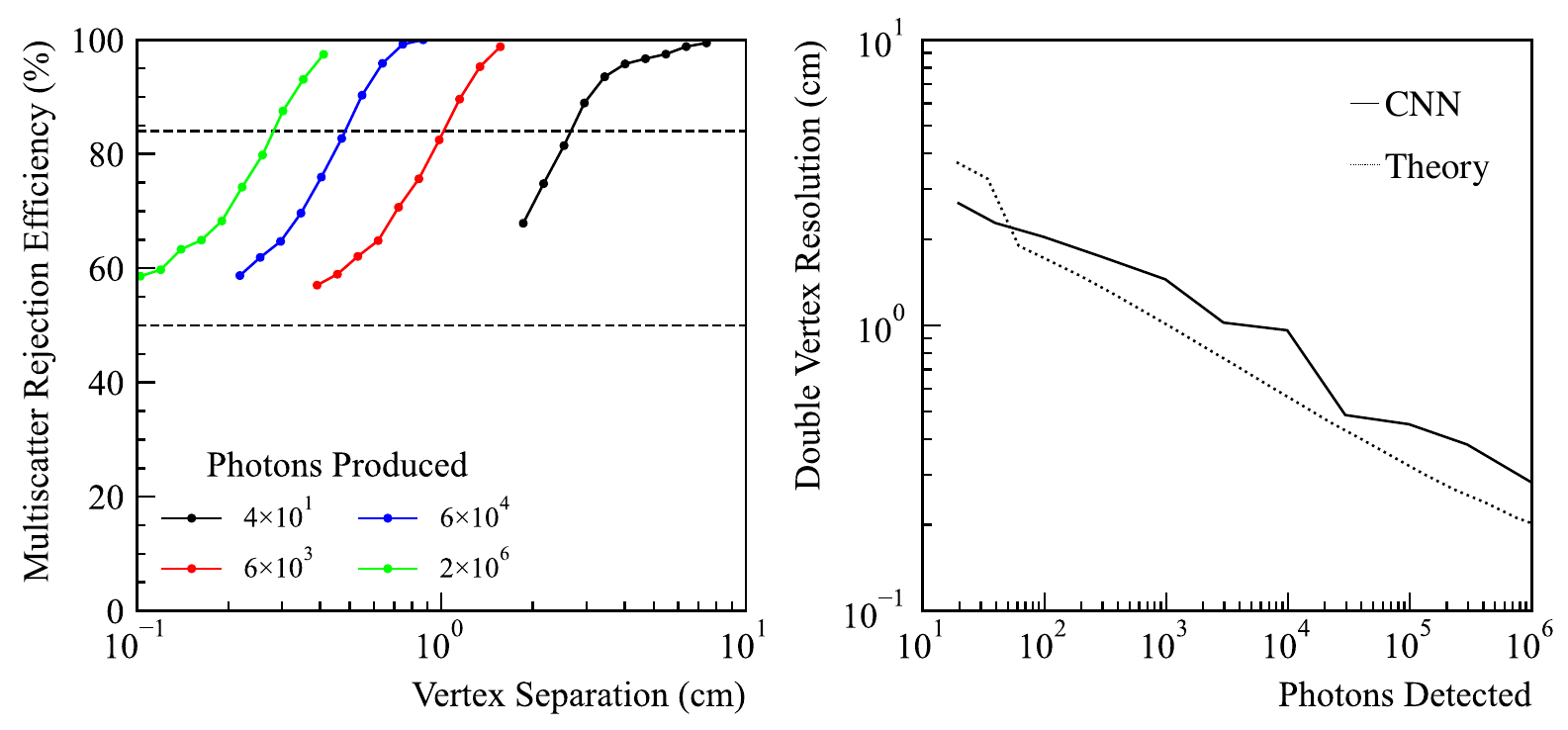}
    \caption{Left - MS rejection efficiency as a function of the vertex separation and photon budget for double vertex events with equal photon splitting. These curves assume 50\% signal acceptance. Right - benchmark MS2 resolution as a function of the number of photons detected using the CNN (solid line) compared to the CRLB (dotted line). We assume the use of 2" PMTs in this study.}
    \label{fig:cnn_performance}
\end{figure}

For a direct comparison with the theoretical limit from FI, CNNs were trained for events with a range of different numbers of photons produced. Simulated SS and MS events not used in the training are then passed to the trained CNNs to assess the classification performance. Following theoretical studies presented in Sec.~\ref{sec:sensitivity}, SS events and the primary vertices of MS events were taken to be located directly under the center of a sensor. We also place the secondary vertices at a separation under investigation but allow a random angle with respect to the primary vertices. At small separations, the two MS vertices are unresolvable and the distribution of CNN scores for SS and MS events overlap, resulting in a $\sim$50\% MS rejection efficiency at 50\% SS acceptance. With increasing separation, CNN performance improves with a sigmoid-like growth until it plateaus near 100\%, as shown in the Fig. \ref{fig:cnn_performance} (left). As in Sec.~\ref{sec:sensitivity}, the benchmark MS resolution is defined as the separation at which 84\% of MS events are rejected (16\% contamination) with 50\% of SS events kept .
The obtained MS2 resolution (with equal photon splitting) from CNN for 2" PMTs is shown in the right panel of Fig. \ref{fig:cnn_performance}, where the CNN result is consistently within 50\% of the theoretical limit. We note that the CNN was trained on a realistic spatial distribution of events but then tested on only those with a vertex directly under a sensor center for comparison with FI predictions to avoid an artificially narrow range of training samples.

\section{Discussion}
\label{sec:disc}

The theoretical and reconstruction studies in Secs.~\ref{sec:theory} and \ref{sec:methods} assume well-controlled MS2 event topologies. 
In a real detector, the stochastic nature of particle interactions causes the interaction energies, event locations, and the separation between vertices to vary continuously. 
In this section we study the capability of a generic TPC (described in Sec.~\ref{sec:detector}) to differentiate physical SS and MS interactions in liquid xenon using optical signals. 
We also discuss how the developed method can help a TPC detector optimize its position reconstruction accuracy and MS discrimination power. 

\subsection{Background rejection in neutrinoless double beta decay experiments}
\label{sec:0vbb}

A key scientific application of SS/MS discrimination in liquid xenon detectors is the search for $0\nu\beta\beta$ decays of \iso{136}{Xe}, for which the signal is the simultaneous emission of two $\beta$ electrons from the same nucleus with total energy summing to $Q = 2.457$~MeV. At these energies, electrons have a typical track length of $\mathcal{O}$(1~mm) in liquid xenon, and therefore the $0\nu\beta\beta$ signals will primarily appear as SS events, with $\sim$10\% that appear as MS events due to the emission of Bremsstrahlung photons from the electron tracks.
The dominant background near the $0\nu\beta\beta$ $Q$-value in TPC-based experiments is the 2.447~MeV gamma ray emitted in the $\beta$ decay of \iso{214}{Bi}, which comes from the primordial $^{238}$U decay chain and is present at trace levels in detector materials~\cite{XLZD2025_0vbb,nEXO_sensitivity_2018}. 
These gammas have a percent-level probability of direct photoabsorption in xenon, which would deposit all the energy at a single site and would not be easily differentiated from $0\nu\beta\beta$ decays. 
However, most of the time the gamma ray will lose a fraction of its energy by Compton scattering at least once before being fully absorbed, and therefore could be identified as MS events. The most important subset of these events are two-scatter events, in which the first scatter deposits the maximum allowed Compton energy, leaving only $\sim$250~keV in the scattered gamma to be photoabsorbed. Since the mean free path of gamma rays decreases with energy, these MS2 events will have a small average separation between vertices, and thus are the primary background pathology. 

\begin{figure}
    \centering
    \includegraphics[width=0.51\textwidth]{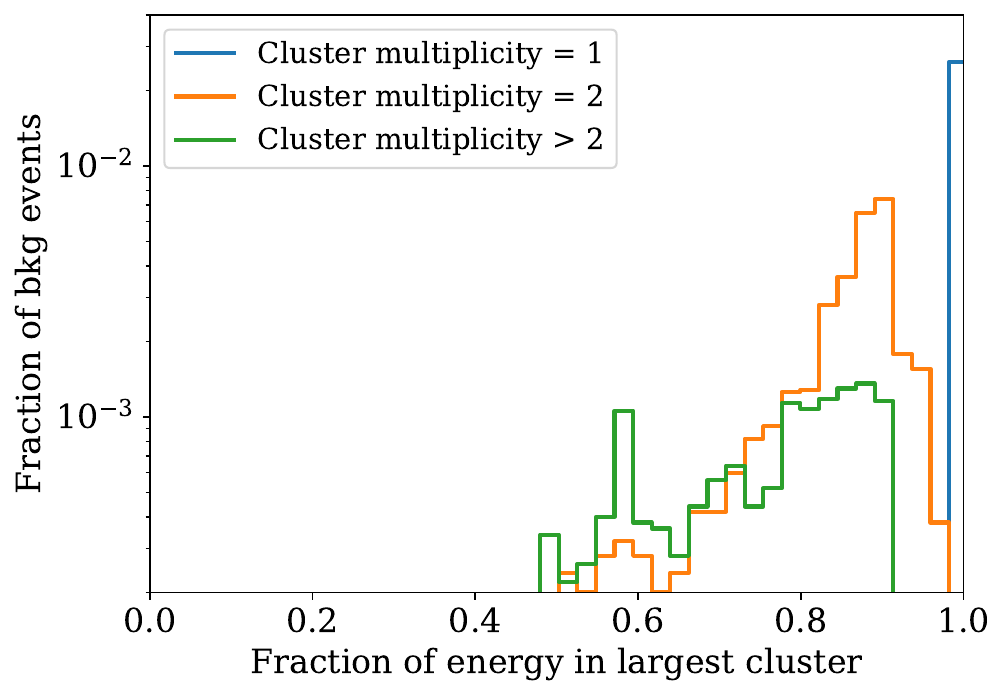}
    \includegraphics[width=0.47\textwidth]{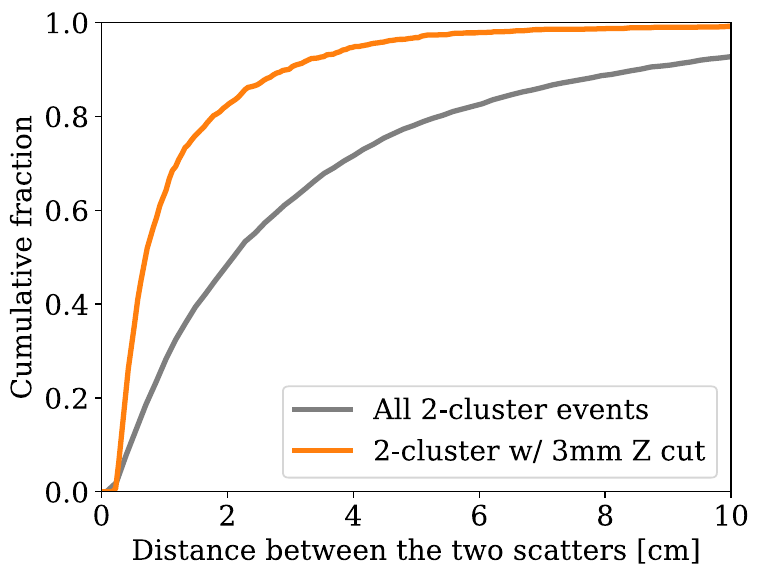}
    \caption{\textbf{left:} the energy fraction contained in the largest cluster from \iso{214}{Bi} gamma-ray interactions in liquid xenon with one, two, and more vertices; the green peak around 0.6 corresponds to pair production by gamma rays. \textbf{right:} the cumulative position distributions between vertices in double-scatter events before and after a 3~mm $z$ cut. The data is obtained from a Monte Carlo simulation with Geant4. }
    \label{fig:bibg}
\end{figure}

We model $^{214}$Bi gamma rays in Geant4 by simulating the isotropic emission of 2.447~MeV gamma rays from a point source in an effectively infinite volume of liquid xenon. The individual energy deposits are clustered into ``interaction vertices'' using the DBSCAN algorithm implemented in \texttt{scikit-learn}~\cite{scikit-learn}, with the distance parameter set to 2~mm. The distance parameter can be varied between 0.5--3~mm with negligible impact on our results. 
After applying a $z$ cut to reject gamma interactions producing energy depositions separated by $>$3~mm in the charge drift direction, $\sim$7\% \iso{214}{Bi} gamma backgrounds remain, comprised of $\sim$2.5\% SS events and $\sim$4.5\% MS events. 
As shown in Figure~\ref{fig:bibg} (left), the MS events are dominated by MS2s, in which the larger vertex preferentially contains $\sim$90\% of the total gamma energy,   
corresponding to the maximum energy that can be transferred to an electron through a single Compton scatter. 
Figure~\ref{fig:bibg} (right) illustrates the spatial separation between the two vertices for double-scatter \iso{214}{Bi} gamma interactions. 
The mean separation is at the cm scale, mainly determined by the photoabsorption cross section of $\sim$250~keV gamma rays in liquid xenon. 
However, the remaining events after a 3~mm $z$ cut have significantly smaller separations between vertices, posing a significant challenge for them to be differentiated from SS signals. 
\iso{136}{Xe} $0\nu\beta\beta$ events are simulated and processed in a similar way. 

At a nominal electric field of 200~V/cm, a 1.23~MeV electron can produce $\sim$60,000 ionization electrons in liquid xenon, leading to $\sim$120,000 total electrons in a single \iso{136}{Xe} $0\nu\beta\beta$ event if cross-track electron-ion recombination~\cite{Xu2025_Recomb} is not considered. 
If we further assume an ionization signal gain of 20 photoelectrons/e- with half of them detected by sensors in the top, a $0\nu\beta\beta$ event could contain $\sim1.2\times10^6$ detectable photoelectrons for $x-y$ position reconstruction, of which up to 300,000 could be detected by a single 2" PMT in our default detector configuration. 
Here we choose a relatively low electron gain to mitigate PMT saturation issues while allowing for single ionization electrons to be identified. 
However, such high levels of light output in a short time window ($\sim 1~\mu s$) can still cause the PMTs to saturate. 
This issue may be further mitigated by reading out the PMT signals from one of the last dynodes instead from the anode, or by optimizing the tapering ratio of the PMT voltage divider. 
It has  been demonstrated that such saturation effects can also be partially corrected in analysis by focusing on the rising edges of the pulses~\cite{XENON:2020iwh}. Saturation leads to an additional uncertainty in the number of photons detected in the affected channel, which could degrade the performance of the SS/MS reconstruction.
The following study ignores such photosensor saturation effects, assuming that either software or hardware solutions can be found to mitigate their impact in the experiment.

Sec.~\ref{sec:sensitivity} estimates a theoretical limit of 7.3~mm for the benchmark MS2 rejection resolution with $1.2\times10^6$ detected photons, assuming the use of 2" PMTs and a 1:9 photon splitting between two vertices. 
This benchmark resolution corresponds to a rejection of 84\% MS2 backgrounds separated by this distance while accepting 50\% SS signals. 
The background rejection power deteriorates for events with smaller separations between vertices or with more imbalanced energy distribution. 
Full calculations of the $0\nu\beta\beta$ signal acceptance and \iso{214}{Bi} background rejection power in our theoretical framework are not straightforward, so we use the MLE and CNN classifier to approximate the TPC's capability in this application. 
As demonstrated in Sec.~\ref{sec:methods}, these methods can approach the theoretical CRLB limits in idealized scenarios. 
Similar to earlier studies, we use an analytical light response function for both sensor hit pattern simulation and reconstruction, and thus ignore uncertainties beyond statistical fluctuations. 

\begin{figure}
    \centering
    \includegraphics[width=0.65\textwidth]{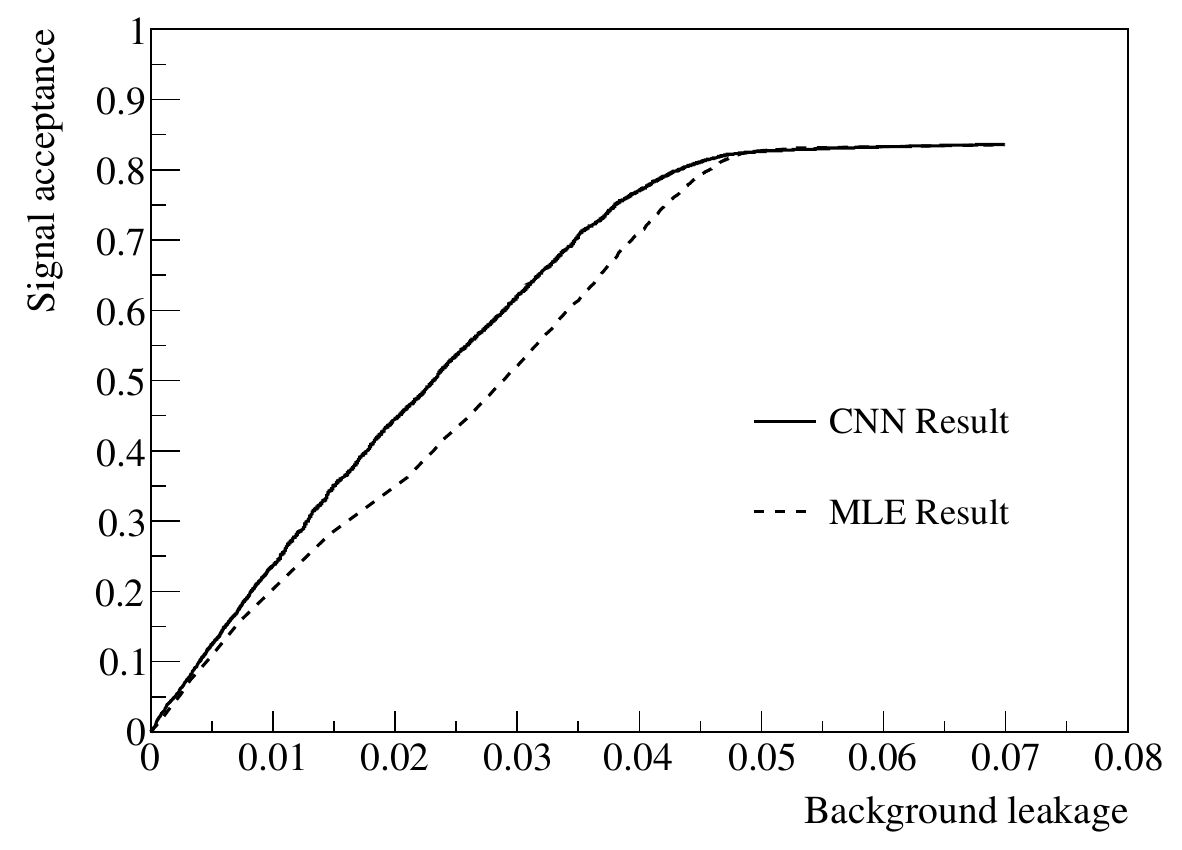}
    \caption{Calculated $0\nu\beta\beta$ signal acceptance and \iso{214}{Bi} background leakage for different cut values, calculated using both the MLE method (dashed line) and the CNN method (solid line) This calculation only considers statistical uncertainties and assumes perfect instrumental response (no noise, no saturation) and reduced smearing in the optical process; detector configuration is illustrated in Fig.~\ref{fig:setup}, with 1~cm electroluminescence region and another 1~cm distance to the sensor array.}
    \label{fig:snr}
\end{figure}

Figure.~\ref{fig:snr} shows the predicted signal acceptance as a function of background leakage for simulated data, using the MLE and CNN methods and the default TPC configuration. 
Here we have already assumed a 3~mm $z$ resolution in the TPC,  which removes $\sim$16\% $0\nu\beta\beta$ signals with bremsstrahlung X-ray emission and  $\sim$93\% of \iso{214}{Bi} backgrounds. 
As a result, the signal acceptance curve saturates around 84\% and background leakage reaches a maximum of $\sim$7\%.  
All events passing the $z$ cut, with multiplicities ranging between 1 and 5, are fed into the simulation framework to produce PMT hit patterns that are then used for discrimination studies. 
The MLE algorithm evaluates the probability for an event to be a MS based on the improvement of fit quality when a second vertex is allowed in the fit, but the CNN is retrained using simulated $0\nu\beta\beta$ and \iso{214}{Bi} events. 
Both algorithms produce similar predictions, with the CNN slightly outperforming the MLE. 
Both methods can reject 1/3 of remaining \iso{214}{Bi} backgrounds with large separations in the $x-y$ plane, with minimal losses of signals. 
Further rejection of background could be achieved at the cost of signal loss; CNN predicts a total background rejection factor of 2 at the cost of losing $\sim$15\% $0\nu\beta\beta$ signals. 
For an actual TPC using only optical readout, however, this performance may not be achieved because our theoretical study assumes no PMT saturation effects and no optical smearing other than Poisson fluctuations, conditions not met in real experiments. 
Nevertheless, this promising result suggests that follow-up investigation into MS reconstruction in optical TPCs, with realistic modeling of detector effects such as photosensor saturation and charge diffusion, are well motivated. 

\subsection{Experiment optimization}
\label{sec:optimize}

Fisher Information is widely used in the design of optical systems where a large parameter space needs to be explored to find the optimal system configuration. 
A performance prediction with FI only requires the system's forward response and its derivative with respect to the parameters of interest, which can often be approximated analytically or numerically. 
It does not require actual data to be acquired with a prototype, and does not need the construction of an estimator to retrieve parameters of interest. 
It is therefore a fast and efficient optimization tool for complex systems at a low cost. 

The CRLB derived from the FI matrix is an estimate of the best reconstruction accuracy for parameters of interest with an unbiased estimator. 
On the one hand, this limit may not be achieved in experiments, especially when simplifications are made in modeling the system behavior (forward model) during sensitivity predictions. 
Under favorable conditions, however, the limits could be reasonably approached; for example, the SS position resolution achieved by the LUX experiment is within 20\% from the predictions made in Sec.~\ref{sec:lux}. 
Further, the comparison of performance for different design options will remain valid if they share similar uncertainty sources. As a result, even when the predicted performance limit could not be reached, FI remains a useful optimization tool. 

On the other hand, a biased estimator, which makes use of a priori knowledge in the estimation of parameters using a specific data set, may exceed the performance predicted by FI. 
For example, an experiment searching for a signal that has been independently and meaningfully constrained may obtain a more stringent measurement on the signal rate than what can be extracted from the experiment itself. 
However, even in these situations, a sensitivity prediction from FI calculations still provides important guidance for the experiment design. 
One may also construct a FI matrix with a likelihood function constrained by a priori knowledge so the optimization process can focus on the most favored signal regions.

A notable result of this work is that the TPC's ability to resolve MS events is significantly worse than that to reconstruct SS events, even when each vertex in the MS event emits the same number of photons as the SS event of reference. 
This is a result of the ambiguity in associating a detected photon in a MS event with a specific vertex. 
The problem is analogous to Rayleigh's criterion in optical microscopy when the Airy patterns of two neighboring objects overlap significantly, leading to a spatial resolution limit near half of the wavelength of the light source. 
In microscopy, super-resolution imaging is achieved by reducing the density of lit objects in the field of view so the MS reconstruction problem is reduced to that for individual single vertices. 
It has been mathematically shown that for objects separated by $>2/f_c$, where $f_c$ is the cutoff frequency of the forward model, the locations of each object, and thus their separation, can be measured with accuracies far exceeding $\lambda$/2~\cite{Candes2014_SuperRes}. 

\begin{figure}
    \centering
    \includegraphics[width=\textwidth]{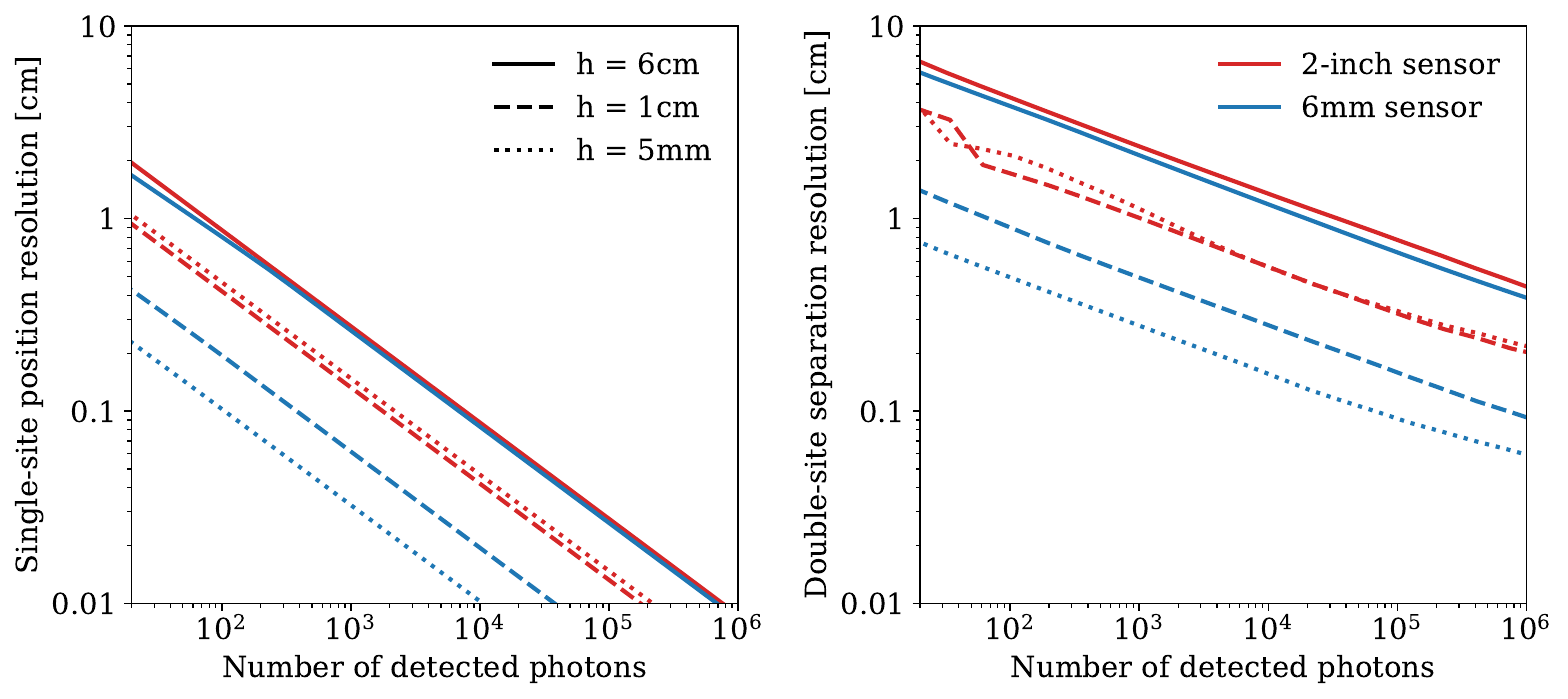}
    \caption{Predicted SS (\textbf{left}) and MS2 (\textbf{right}) reconstruction resolutions for 2" photosensors (red lines) and 6~mm photosensors (blue lines) with different distances between the electroluminescence region and the photosensor array (denoted as $h$): 6~cm (solid lines), 1~cm (dashed lines), 0.5~cm (dotted lines). For MS2 events a 1:1 energy splitting between the two sites is assumed. }
    \label{fig:comp1cm}
\end{figure}

In our TPC model, the separation between MS event vertices is governed by physical laws and cannot be increased, while the observed spread of the light response function in a TPC can be strongly shaped by choices of detector configuration. 
Smaller sensor pixels can lead to sharper light response functions and better position resolution, as shown in Figs.~\ref{fig:LRF} and \ref{fig:dresn}. 
For sensor sizes comparable to, or smaller than the distance between the photon source and the sensor plane, the gain from reducing sensor size diminishes. 
Therefore, the benefits of small sensors can only be realized by choosing a small physical separation between the light source and the sensor plane. 
This is demonstrated in Fig.~\ref{fig:comp1cm}, where we compare the predicted SS (left) and MS2 (right, assuming 1:1 photon splitting between two vertices) resolutions for varying distances between the electroluminescence region and the photosensor plane, assuming the use of 2" sensors (red lines) and 6~mm sensors (blue lines). 
Compared to the baseline model of 1~cm sensor separation (dashed lines), a 0.5~cm separation~\footnote{For the 0.5~cm sensor separation study, the electroluminescence region thickness is also reduced to 0.5~cm. } (dotted lines) improves the resolution for 6~mm sensors by approximately a factor of 2; the results for 2" sensors remain largely unchanged because the photon distribution is dominated by the large size of the sensor in these two cases. 
When the sensor separation from the electroluminescence region is increased from 1~cm to 6~cm (solid lines), which was the configuration of the LUX detector, both SS and MS resolutions substantially deteriorate. In this configuration, the 2" sensor results and the 6~mm sensor results are nearly identical, because the photon distributions in both scenarios are dominated by the large light source-sensor separation. 

The theoretical gain predicted for small sensor sizes and small light source-sensor distances is accompanied by increased costs of pixilated sensors, large numbers of readout electronics and more demanding high voltage requirements. In addition, the sensor saturation problem may also deteriorate in such configurations.   
Such practical constraints shall be considered in the optimization of the detector configuration for desired position reconstruction and MS discrimination performance. 
This is a potential application of the method outlined in this work. 

\section{Conclusion}
\label{sec:conclusion}

We use Fisher Information to predict the single-site and double-site event discrimination power of a generic idealized dual-phase TPC. 
The theoretical study is carried out for TPC models using sensors of different sizes, and the results are presented as functions of event location, energy and inter-vertex separation for double-vertex events. 
We further conduct event reconstruction tests with simulated data using the principal-component analysis, the maximum likelihood estimator and a convolutional neural network. 
The results obtained from reconstruction efforts approach the theoretical predictions by Fisher Information in all scenarios studied, confirming the applicability of this method to calculating position reconstruction in TPCs. 
The results are then used to study the multi-site background rejection in TPC-based neutrinoless double beta decay experiments. We also discuss how Fisher Information may be used to optimize future experiments for improved performance.

\input{Acknowledgements}

\appendix

\section{Fisher Information for independent Poisson distributions}
\label{sec:emlfi}

The general definition of a likelihood for $K$ independent Poisson measurements is:
\begin{equation}
    \mathcal{L} (n_1...n_K|\Theta) = \prod_{k}^K \frac{\eta_k^{n_k} e^{-\eta_k}}{n_k!}
\end{equation}
where $\Theta$ is a set of parameters and $n_k$ is the observed count in measurement $k$ with an expectation of $\eta_k(\Theta)$. The log-likelihood is simply
\begin{equation}
    \log\left(\mathcal{L}(n_1...n_K|\Theta)\right) =\sum_k \left(n_k \log(\eta_k) -\eta_k - \log (n_k!)\right)
\end{equation}
The $\{i,j\}$th element of the Fisher Information Matrix is defined as 
\begin{equation}
    \mathcal{I}(\Theta)_{i,j} = E\left( \frac{\partial \log(\mathcal{L})}{\partial \theta_i} \frac{\partial \log(\mathcal{L})}{\partial \theta_j}\right)
\end{equation}
where the expectation is calculated over all observable space - in our case it is over all possible $n_k$ values.
First let's calculate the derivative of $\mathcal{L}$ with respect to $\theta_i$
\begin{equation}
    \frac{\partial \log(\mathcal{L})}{\partial \theta_i} = \sum_k \left(\frac{n_k}{\eta_k}\frac{\partial \eta_k}{\partial \theta_i}-\frac{\partial \eta_k}{\partial \theta_i} \right)
    = \sum_k \frac{n_k-\eta_k}{\eta_k}\frac{\partial \eta_k}{\partial \theta_i}
    \label{eq:score}
\end{equation}
So the Fisher Information Matrix element becomes
\begin{align}
        \mathcal{I}(\Theta)_{i,j} & = E\left( \left( \sum_k \frac{n_k-\eta_k}{\eta_k}\frac{\partial \eta_k}{\partial \theta_i}\right)  \left( \sum_{k'} \frac{n_{k'}-\eta_{k'}}{\eta_{k'}}\frac{\partial \eta_{k'}}{\partial \theta_j}\right)\right) \nonumber \\
        &=\sum_{k,k'} \frac{\partial \eta_k}{\partial \theta_i} \frac{\partial \eta_{k'}}{\partial \theta_j}E\left( \frac{n_k-\eta_k}{\eta_k} \frac{n_{k'}-\eta_{k'}}{\eta_{k'}}\right) \nonumber \\
        &=\sum_{k} \frac{\partial \eta_k}{\partial \theta_i} \frac{\partial \eta_{k}}{\partial \theta_j}E\left( \frac{n_k-\eta_k}{\eta_k} \right)^2 +\sum_{k,k'\neq k} \frac{\partial \eta_k}{\partial \theta_i} \frac{\partial \eta_{k'}}{\partial \theta_j}E\left( \frac{n_k-\eta_k}{\eta_k} \frac{n_{k'}-\eta_{k'}}{\eta_{k'}}\right)
\end{align}
Because $n_k$ and $n_{k'}$ are independent when $k'\neq k$, their expectations can be calculated separately. 
\begin{equation}
    E\left( \frac{n_k-\eta_k}{\eta_k} \frac{n_{k'}-\eta_{k'}}{\eta_{k'}}\right) = E\left( \frac{n_k-\eta_k}{\eta_k} \right) E\left( \frac{n_{k'}-\eta_{k'}}{\eta_{k'}}\right)=0
\end{equation}
This term vanishes because $E(n_k)=\eta_k$ for a Poisson distribution. Using $E(n_k^2)=\eta_k^2+\eta_k$ for a Poissonian process, we obtain the final result as
\begin{equation}
    \mathcal{I}(\Theta)_{i,j}  = \sum_{k} \frac{\partial \eta_k}{\partial \theta_i} \frac{\partial \eta_{k}}{\partial \theta_j} \frac{1}{\eta_k}
\label{eq:fielem}
\end{equation}
which is the standard Fisher Information Matrix for an independent Poissonian process.

Alternatively, the FI matrix may be constructed from
\begin{equation}
    \mathcal{I}(\Theta)_{i,j} = - E\left( \frac{\partial }{\partial \theta_i} \frac{\partial }{\partial \theta_j} \log(\mathcal{L})\right)
\end{equation}
Again we can start with Eq.~\ref{eq:score}
\begin{equation}
     \frac{\partial }{\partial \theta_i} \frac{\partial }{\partial \theta_j} \log(\mathcal{L}) = \sum_k \frac{\partial }{\partial \theta_i} \left( \frac{n_k-\eta_k}{\eta_k}\frac{\partial \eta_k}{\partial \theta_j}\right) = \sum_k\left(-\frac{n_k}{\eta_k^2}\frac{\partial \eta_k}{\partial \theta_i} \frac{\partial \eta_{k}}{\partial \theta_j}+\frac{n_k-\eta_k}{\eta_k}\frac{\partial^2 \eta_k}{\partial \theta_i \partial \theta_j}\right)
\end{equation}
Using $E(n_k)=\eta_k$, we can easily come to the same FI matrix element as in Eq.~\ref{eq:fielem}
\begin{equation}
    \mathcal{I}(\Theta)_{i,j} = - E\left( \frac{\partial }{\partial \theta_i} \frac{\partial }{\partial \theta_j} \log(\mathcal{L})\right) = \sum_{k} \frac{\partial \eta_k}{\partial \theta_i} \frac{\partial \eta_{k}}{\partial \theta_j} \frac{1}{\eta_k}
\end{equation}

\bibliography{main.bib}{}
\bibliographystyle{ieeetr}
\end{document}

%% file: Acknowledgements.tex
\section*{Acknowledgments}

This work was performed under the auspices of the U.S. Department of Energy by Lawrence Livermore National Laboratory (LLNL) under Contract DE-AC52-07NA27344. 
It is supported by the LLNL-LDRD Program under Project No. 25-LW-034 and by the U.S. Department of Energy (DOE) Office of Science, Office of High Energy Physics under Work Proposal Number SCW1676 awarded to LLNL. 

LLNL IM release number: LLNL-JRNL-2011380. 